\PassOptionsToPackage{bottom}{footmisc}
\documentclass{technicalreport}
\usepackage{footmisc}
\renewcommand{\reporttitle}{MMIntentFlow: An Intent-Driven Real-Time  Multimodal Search System with Image–Text Queries at Xiaohongshu}
\renewcommand{\reportsubtitle}{}
\renewcommand{\reportauthors}{%
Dongjia Zhao\textsuperscript{*}, Yuchen Ma\textsuperscript{*}, Chunsheng Shi\textsuperscript{*}, Ziqiu Gui\\
Chao Yuan, Jianing Zhu, Zhaoxing Zhang, Yahui Liu\\
Daoxin Zhang, Yi Wu, Yao Hu, Cheng Chen\textsuperscript{\dag}}
\renewcommand{\reportaffiliation}{Xiaohongshu Inc}
\renewcommand{\reportversion}{}
\renewcommand{\reportdate}{}
\renewcommand{\reportshorttitle}{MMIntentFlow}
\definecolor{reportaccent}{RGB}{220,38,38}

\usepackage{mathtools}
\usepackage{xspace,multirow,float,adjustbox}
\usepackage{ragged2e,placeins}
\usepackage{natbib}
\usepackage{tocloft}
\usepackage{xltabular}
\hypersetup{
    colorlinks=true,
    citecolor=blue!55!black,
    linkcolor=reportaccent!80!black,
    urlcolor=blue!65!black,
    pdftitle={\reporttitle},
    pdfauthor={Dongjia Zhao, Yuchen Ma, Chunsheng Shi, Ziqiu Gui, Chao Yuan, Jianing Zhu, Zhaoxing Zhang, Yahui Liu, Daoxin Zhang, Yi Wu, Yao Hu, Cheng Chen},
    pdfsubject={MMIntentFlow Technical Report}
}

\providecommand{\Description}[1]{}
\newcommand{\method}{\textsc{MMIntentFlow}\xspace}

\newcolumntype{Y}{>{\raggedright\arraybackslash}X}
\newcommand{\retrievalmetric}[1]{\fontsize{7.5}{9}\selectfont\bfseries #1}
\newcommand{\reporttablefont}{\fontsize{9}{11}\selectfont\setlength{\tabcolsep}{5pt}\renewcommand{\arraystretch}{1.2}}
\newcommand{\reporttablehead}[1]{\textbf{#1}}
\newcommand{\reporttablegroup}[2]{\addlinespace[3pt]\multicolumn{#1}{@{}l}{\textbf{#2}}\\[1.5pt]}

\begin{document}
\begin{abstract}
Multimodal search with image–text queries helps users find notes in user-generated content (UGC) communities and serves as a key search tool for multimodal search agents. For such queries, query–note relevance depends on the user’s retrieval intent and cannot be determined by visual or topical similarity alone. Simply reranking visual search results cannot fully address this challenge, as notes relevant to the user’s intent may be missing from the initial candidate set. This motivates explicit intent modeling to align candidate retrieval and reranking under consistent relevance criteria. We present \method, an intent-driven real-time multimodal search system deployed at Xiaohongshu over an index on the order of ten billion entries. \method represents retrieval objectives with a hierarchical intent taxonomy and uses intent-specific relevance criteria to guide data construction, representation learning, and reranking. An offline agentic data workflow acquires candidates and verifies their relevance under these criteria to produce training supervision. The embedding model combines intent-aware contrastive learning, graded ranking, and distribution distillation, while a compact reranker assesses query--candidate evidence conditioned on predicted hierarchical intents. Evaluations demonstrate that \method delivers substantial gains in search engagement while supporting real-time search at industrial scale.
In a user-level A/B test on the Xiaohongshu platform, \method outperforms the existing production system based on visual search followed by reranking, achieving relative gains of 28.17\% in Top-4 query click-through rate, 18.96\% in query click-through rate, and 12.56\% in average time spent on search results pages per user.
Production monitoring records an end-to-end P50 latency of 1.07~s.
\end{abstract}
\maketitle
\begingroup
\renewcommand{\thefootnote}{\fnsymbol{footnote}}
\footnotetext[1]{Equal contribution.}
\footnotetext[2]{Project lead \& corresponding author: \href{mailto:chengchen7@xiaohongshu.com}{\nolinkurl{chengchen7@xiaohongshu.com}}}
\endgroup
\par\vspace{12pt}
\noindent\textbf{Keywords:} multimodal search; user-generated content; hierarchical intent modeling; agentic data construction; multimodal representation learning; intent-conditioned reranking; real-time search.

\clearpage
\tableofcontents
\clearpage

\section{Introduction}

\begin{figure}[H]
\centering
\includegraphics[width=\linewidth]{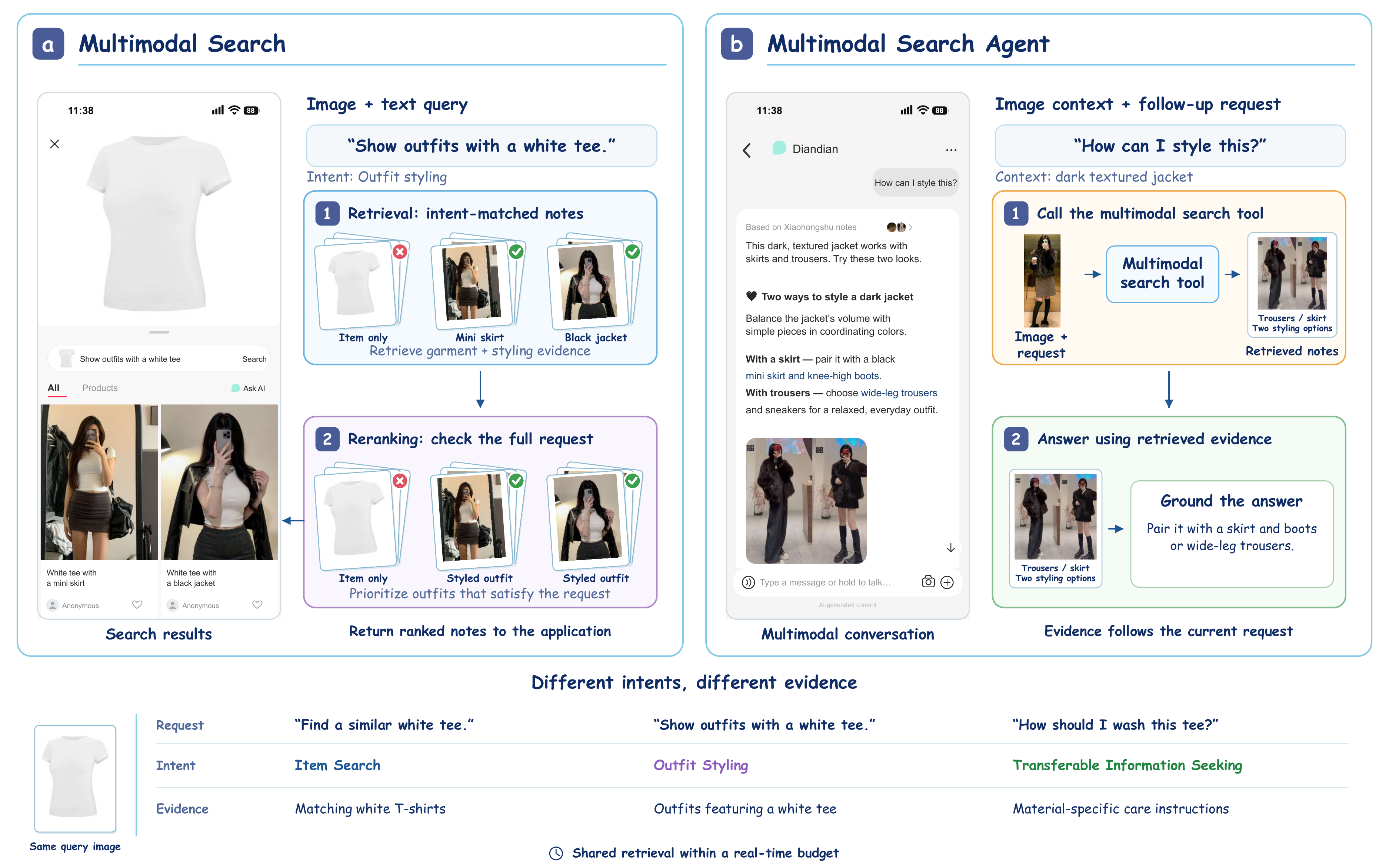}
\caption{Two applications of multimodal search. (a) Image--text retrieval and reranking return outfit notes for a white T-shirt. (b) A multimodal search agent uses retrieved outfit examples to support styling suggestions for a dark jacket. The application interfaces and responses are reconstructed for illustration. The bottom comparison links three requests about the same T-shirt image to their retrieval intents and required evidence.}
\Description{Two panels show Multimodal Search and a Multimodal Search Agent. On the left, a white T-shirt product image and a styling request are used to retrieve and rerank notes. A product-only image lacks the requested styling evidence, whereas two relevant examples show a white T-shirt with a mini skirt and with a black jacket. On the right, an elevator photograph supplies the visual context for a styling question about a dark jacket. A retrieved side-by-side photograph shows wide-leg trousers with sneakers and a mini skirt with knee-high boots. The reconstructed Diandian interface presents illustrative styling suggestions based on these examples. Below, three requests about the same T-shirt image illustrate Item Search, Outfit Styling, and Transferable Information Seeking.}
\label{fig:applications}
\end{figure}
\nopagebreak[4]

User-generated content (UGC) communities enable users to share and discover visual examples, personal experiences, and practical knowledge. At Xiaohongshu, this content is organized into image--text notes, allowing a shared corpus to support both visual retrieval and information seeking. Users combine a reference image with a textual request to retrieve these notes, and the same retrieval service supplies evidence to multimodal search agents (Figure~\ref{fig:applications}). Serving these applications requires identifying useful content from an index on the order of ten billion entries within the response-time constraints of online interaction.

Within this UGC corpus, notes about the same object may depict its appearance, report experiences with it, or provide practical guidance. For example, the same T-shirt image may accompany a request to find a similar item, explore outfit combinations, or learn how to wash the garment. A note showing a visually matching T-shirt may satisfy the first request without providing either a styling example or applicable care instructions. Conversely, a care guide illustrated with another garment may still be useful when its material and care requirements apply. User intent therefore determines both the required correspondence to the reference image and the evidence needed to satisfy the request.

Recent work has extended retrieval to settings in which queries and candidates can both contain images and text. UniIR~\citep{wei2024uniir} uses instruction-guided training to support multiple modality combinations, including image--text-to-image--text retrieval in M-BEIR. WeMM-Embedding~\citep{zhou2026wemm} learns representations for mixed-modal queries and candidates using graded relevance supervision and knowledge transfer. UMER~\citep{chen2026umer} combines embedding retrieval and ranking through pair-aware discriminative reasoning and mutual distillation. These advances improve modality coverage and candidate discrimination, but UGC search additionally requires a relevance definition grounded in structured intent for recurrent retrieval objectives. A candidate that is a valid positive for one objective may fail a decisive requirement under another; moreover, useful content omitted during retrieval cannot be recovered through reranking. The system-level challenge is therefore to make these objective-dependent distinctions explicit and apply them consistently across training-data construction, candidate retrieval, and relevance assessment.

We introduce \method, the first real-time multimodal search system, to our knowledge, to use explicit structured retrieval intents to guide data construction, representation learning, and reranking for image--text queries in a real-world UGC community. The system uses a retrieval-objective hierarchy that describes the result sought, rather than the category of the depicted object. For actionable requests, the first level distinguishes Visual Search, where visual content provides the requested result, from Information Seeking, where textual information is additionally required. The second level specifies the retrieval objective within each group. Combined with the target and constraints in the original query, these intents define the required correspondence and evidence for judging relevance. This formulation makes intent a common basis for learning what to retrieve and determining whether retrieved content satisfies the request.

To translate this intent formulation into retrieval behavior, we develop an agentic data workflow that constructs and verifies supervision for both retrieval stages. The workflow acquires candidate notes, gathers missing evidence, and checks whether each candidate satisfies the intent-specific requirements. Verified positives and hard negatives train the embedding to distinguish satisfying candidates from confusable alternatives; graded relevance supervision and offline teacher distributions further refine candidate ordering. The reranker uses the same relevance criteria to assess the original query and candidate evidence, conditioned on first- and second-level intents predicted by a compact recognizer. Intent thus governs both the relationships learned by the embedding and the relevance decisions made after retrieval.

Deploying this design at industrial scale requires controlling both online computation and the cost of updating the document index. We verify training evidence and obtain teacher supervision offline, then train compact models for online retrieval and relevance prediction. During serving, vector search retrieves candidates from a precomputed document index. The reranker then applies the intent-specific criteria to these candidates. To support model iteration without corpus-wide re-encoding, query-side feature adaptation and residual projection further refine retrieval within the fixed document space.

We evaluate the design through offline experiments and production deployment. Offline comparisons examine intent-aware and generic supervision under the same training procedure. In an online A/B test with user-level traffic allocation, the complete system increases Top-4 query click-through rate by 28.17\%, query click-through rate by 18.96\%, and average time spent on search results pages per user by 12.56\% relative to the production control. Separate service monitoring records end-to-end request latencies of 1.07~s at P50 and 1.34~s at P90. The main contributions of this report are summarized as follows:
\begin{itemize}
\item \textbf{Structured intent modeling for UGC retrieval.} We introduce a retrieval-objective hierarchy for multimodal search in UGC communities and derive intent-specific relevance criteria that guide training-data construction, embedding-based candidate retrieval, and relevance reranking.
\item \textbf{Intent-driven agentic data construction, representation learning, and reranking.} We develop an agentic workflow for candidate acquisition and evidence verification. We combine the resulting intent-aware contrastive and graded-ranking supervision with distribution distillation for representation learning, and train a reranker to assess candidate evidence under explicit hierarchical intents.
\item \textbf{Real-time industrial deployment and evaluation.} We deploy the complete intent-driven pipeline at Xiaohongshu and evaluate its benefits through controlled offline supervision comparisons and a production A/B test. The experiments quantify retrieval and engagement gains alongside end-to-end serving latency at industrial scale.
\end{itemize}

\section{Related Work}
\paragraph{Multimodal retrieval and information seeking.}
Multimodal retrieval supports different objectives, from locating a visual target to finding information relevant to a request. TIRG~\citep{vo2019composing}, FashionIQ~\citep{wu2021fashioniq}, and CIRR~\citep{liu2021cirr} study image retrieval conditioned on a reference image and textual modifications. Image-copy detection~\citep{douze2021imagesimilarity} and OutfitTransformer~\citep{sarkar2023outfit} address the identification of image copies and the retrieval of compatible fashion items, respectively. For information seeking, PreFLMR~\citep{lin2024preflmr} retrieves knowledge for visual questions, while MuKA~\citep{deng2025muka} requires positive multimodal documents to match the queried entity and contain the answer. Relevance can also depend on information that transfers across examples: the theorem-based tasks in BRIGHT~\citep{su2025bright} retrieve solved problems sharing an underlying theorem, and MM-BRIGHT~\citep{abdallah2026mmbright} distinguishes reasoning-supporting evidence from topically similar but insufficient documents.

UniIR~\citep{wei2024uniir}, VLM2Vec~\citep{jiang2025vlm2vec}, and GME~\citep{zhang2025gme} unify multiple tasks and modality combinations through instruction-guided representations. Qilin~\citep{chen2025qilin} provides a community-search dataset with multimodal notes and user feedback. \method studies visual retrieval and information seeking within a shared UGC note corpus, where the retrieval objective determines the required target correspondence and the applicability of a note's information.

\paragraph{Intent-aware retrieval and data construction.}
The dependence of relevance on information needs and user situations is well established~\citep{borlund2003relevance}. I3~\citep{pan2024i3} incorporates instructions through intent introspection, and FollowIR~\citep{weller2025followir} evaluates how retrieval models respond when revised assessment instructions change document relevance. Such conditions also shape training supervision. TART~\citep{asai2023tart} introduces instruction-unfollowing negatives and uses an instruction-aware reranker to refine positive and hard-negative sets. Promptriever~\citep{weller2025promptriever} generates passages that are relevant to the query but become non-relevant under an additional instruction. For multimodal retrieval, MM-Embed~\citep{lin2025mmembed} uses modality-aware hard-negative mining to mitigate modality bias.

Synthetic supervision broadens the range of retrieval instructions. MagicLens~\citep{zhang2024magiclens} generates open-ended instructions for image pairs mined from the same webpage, while MegaPairs~\citep{zhou2025megapairs} mines diverse pairs from open-domain image corpora and generates corresponding instructions. mmE5~\citep{chen2025mme5} combines visual interpretation with instruction and text synthesis, followed by self-evaluation and revision within a single generation pass. Agent-based curation also appears in industrial retrieval: Pailitao-VL~\citep{chen2026pailitaovl} uses a Plan--Propose--Organize--Review workflow to curate an instance-identity prototype library. \method uses an agentic workflow to verify intent-specific requirements for UGC notes, including target correspondence, requested relations, and information applicability, when constructing positives, hard negatives, and relevance labels.

\paragraph{Multimodal representation learning and relevance modeling.}
DME~\citep{chen2026dme} augments contrastive learning with evidence-grounded typed latent reasoning and cross-conditional reconstruction, grounding representations in local evidence and preserving semantics from the paired query or document. Qwen3-VL-Embedding~\citep{li2026qwen3vl} uses reranker scores to supervise representation learning. WeMM-Embedding~\citep{zhou2026wemm} combines graded relevance learning with reranker supervision on selected tasks and distillation from a larger embedding model. UMER~\citep{chen2026umer} jointly learns retrieval representations and ranking within a single multimodal model through pair-aware reasoning, verified triplets, and mutual distillation. Its reasoning explicitly describes query intent and examines supporting or contradictory candidate evidence.

For candidate reranking, LamRA-Rank~\citep{liu2025lamra} supports pointwise binary relevance prediction and listwise target-position prediction. In retrieval evaluation, graded judgments capture degrees of relevance beyond binary labels~\citep{jarvelin2002cumulated}. Pailitao-VL~\citep{chen2026pailitaovl} combines comparative ranking with a four-level scale spanning instance, concept, functional, and irrelevant matches. In \method, a discrete hierarchy makes recurring retrieval objectives explicit. Under each objective, relevance grades distinguish content that satisfies the request from content that is merely related or irrelevant. These intent-specific criteria guide both embedding supervision and intent-conditioned reranker training.

\section{Intent-Conditioned Relevance}
\label{sec:relevance}
\subsection{Task and Intent Hierarchy}
\label{sec:intent-scope}
We study region-level multimodal retrieval over user-generated notes. A query $q=(x_q,t_q,b_q)$ consists of a reference image $x_q$, query text $t_q$, and a bounding box $b_q$ specifying the query region. The service initializes $b_q$ using an existing object detector; users may adjust the default box or define their own region, including the full image. On the document side, a retrieval unit $d=(x_d,t_d,b_d)$ comprises a note image $x_d$, the note's title and body $t_d$, and a candidate-region box $b_d$ produced by the pipeline's object detector. Let $\mathcal D$ denote the collection of these region-associated document units. Query--document matching compares the visual content designated by $b_q$ and $b_d$ together with the associated text and image context. Each document unit remains linked to its source note, which provides the content returned to the user.

The intent hierarchy is organized by the requested outcome and the evidence needed to establish relevance. Requests about the same object may seek different results, while the same retrieval objective can apply across object categories. The first level groups requests by shared evidence requirements; the second distinguishes objectives with different matching conditions. These distinctions guide both the construction of training supervision and the assessment of candidate relevance.

\begin{figure}[H]
\centering
\includegraphics[width=\linewidth]{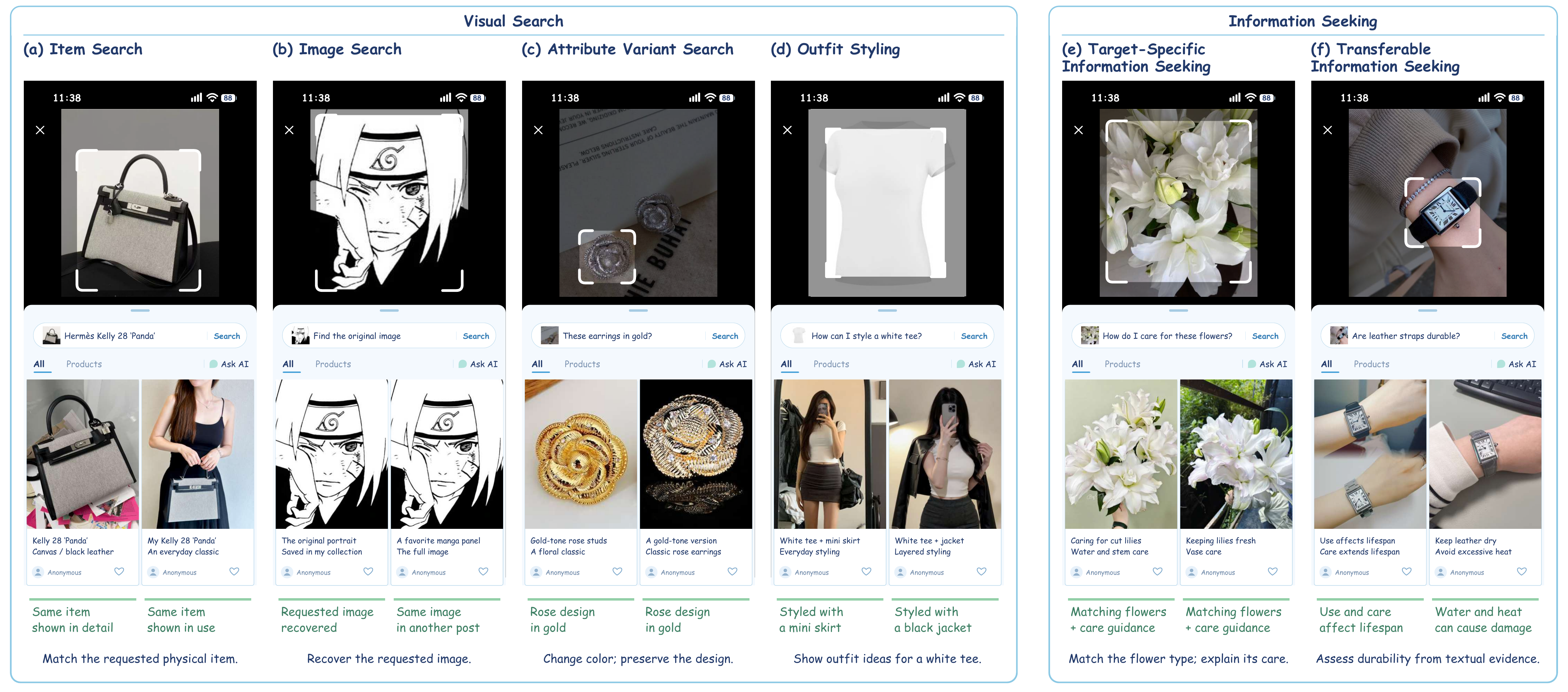}
\caption{Intent-dependent evidence requirements for six actionable retrieval intents. Reconstructed search panels illustrate (a) Item Search, (b) Image Search, (c) Attribute Variant Search, (d) Outfit Styling, (e) Target-Specific Information Seeking, and (f) Transferable Information Seeking. Each panel pairs a query with two illustrative relevant notes. White corner brackets mark the query region; the surrounding image is dimmed.}
\Description{Six search panels are arranged in a single row. Panels (a--d) illustrate Visual Search, and panels (e--f) illustrate Information Seeking. The visual examples show the requested item, the original image, a color variant, and outfits pairing a white T-shirt with a mini skirt or a black jacket. The information-seeking examples show care guidance for lilies and textual information about leather-strap durability. The latter concerns the effects of use, care, moisture, and heat rather than the identity of a particular watch. Each query image contains white corner brackets marking a local target or the full image, with the area outside the brackets dimmed. Each panel contains two illustrative relevant notes.}
\label{fig:motivation}
\end{figure}
\nopagebreak[4]

The hierarchy comprises seven second-level intents under three first-level intents: Visual Search (Figure~\ref{fig:motivation}(a--d)), Information Seeking (Figure~\ref{fig:motivation}(e--f)), and Unclear Intent. For query $q$, let $i_q^{(1)}\in\mathcal I^{(1)}$ denote its first-level intent and $i_q^{(2)}\in\mathcal I^{(2)}(i_q^{(1)})$ its second-level intent.

For actionable requests, \textbf{Visual Search} seeks a result conveyed by the candidate's visual content, such as an item, an image, or an outfit combination. \textbf{Information Seeking} additionally requires the candidate's text to answer a question or provide an explanation or practical guidance. Both groups retrieve multimodal notes; they differ in the evidence needed to satisfy the request.

Within Visual Search, the second level distinguishes four types of visual result.

\textbf{Item Search} seeks notes depicting the requested physical item or an alternative permitted by the query. For exact-item requests, a shared category or similar appearance does not establish a match.

\textbf{Image Search} seeks the visual asset itself, such as an original image or a version satisfying specified source or resolution requirements.

\Needspace{4\baselineskip}
\textbf{Attribute Variant Search} requires a specified attribute change while preserving other query-relevant characteristics.

\textbf{Outfit Styling} seeks combinations or styling examples for the reference garment or a stated styling goal. The result must convey the requested combination or styling relation.

Within Information Seeking, the second level distinguishes whether the answer must concern the designated target.

\textbf{Target-Specific Information Seeking} requires information about that target: care guidance for the depicted lilies, for example, must apply to that flower type.

\textbf{Transferable Information Seeking} permits information from a different object or case, provided that it answers the question under applicable conditions. The supporting image need not depict the same or a visually similar object. For leather-strap durability, relevant experience may transfer across watch models when material, use, and care conditions make it applicable.

\textbf{Unclear Intent} applies when the image and text jointly fail to establish an actionable retrieval objective; it is represented at both levels of the hierarchy. Short or unfamiliar text alone is insufficient to assign this label. Appendix~\ref{app:intent-taxonomy} provides the complete definitions.

The hierarchy specifies the type of result sought, while the complete query determines the target and constraints that a candidate must satisfy. In the following, $i$ abbreviates the second-level intent $i_q^{(2)}$ when the query is implicit. Section~\ref{sec:relevance-criteria} defines the corresponding relevance grades.

\FloatBarrier

\subsection{Relevance Criteria and Graded Judgments}
\label{sec:relevance-criteria}

\begin{figure}[H]
\centering
\includegraphics[width=\linewidth]{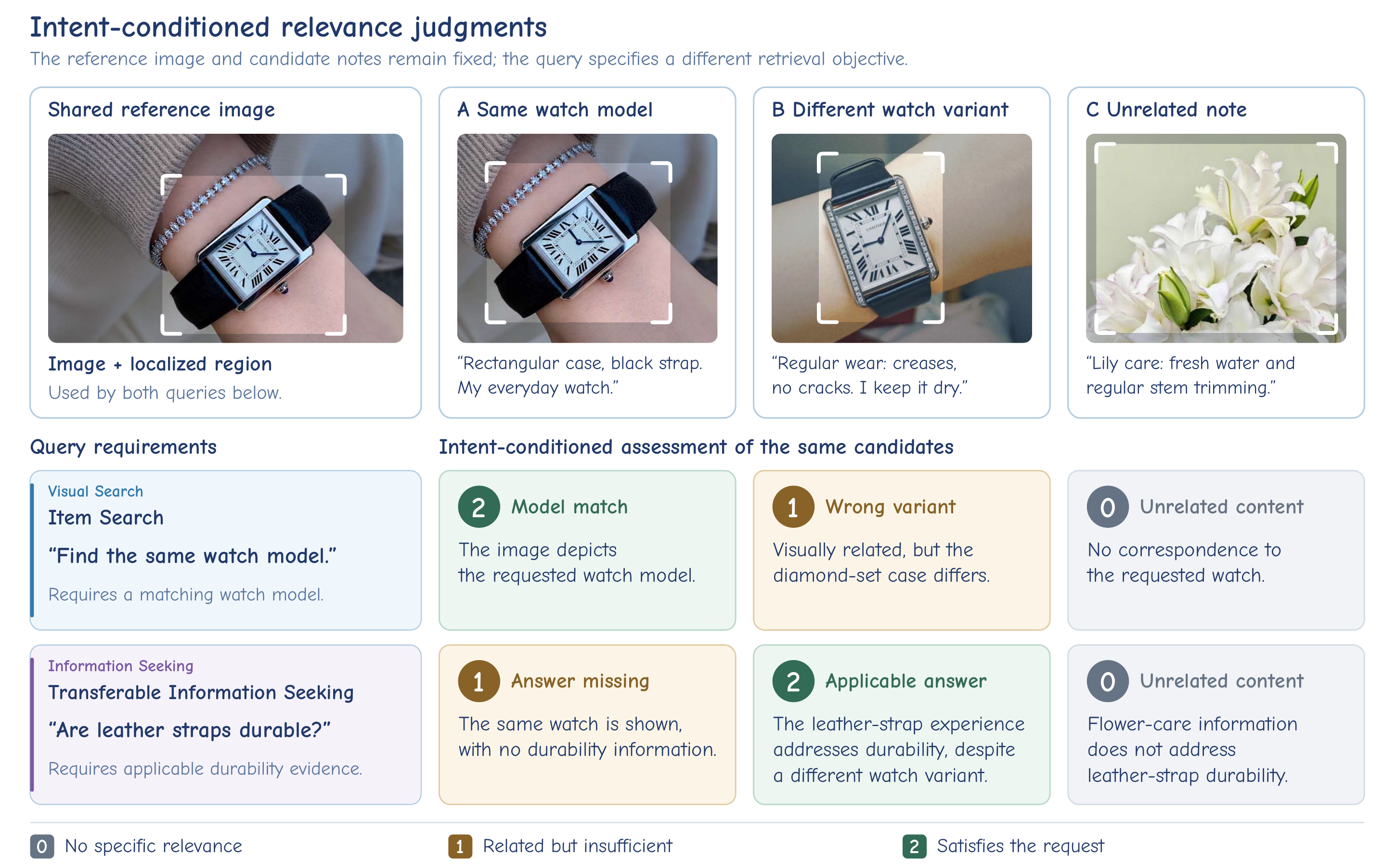}
\caption{Intent-conditioned relevance judgments for two queries sharing a reference image and candidate set. Item Search requires the requested watch model, whereas Transferable Information Seeking requires applicable durability information. The same candidates therefore receive different grades under the two requests. White corner brackets denote query and document regions; note excerpts are illustrative.}
\Description{A shared watch reference image is paired with two queries and three candidate notes. Candidate A shows the same watch model with an appearance description. Candidate B shows a different, diamond-set variant and describes creasing without cracks during regular wear while keeping the leather strap dry. Candidate C concerns lily care. For Find the same watch model, the candidates receive grades 2, 1, and 0. For Are leather straps durable, they receive grades 1, 2, and 0. Each grade is accompanied by a reason. White corner brackets identify the query and candidate regions.}
\label{fig:relevance-judgments}
\end{figure}
\nopagebreak[4]

Figure~\ref{fig:relevance-judgments} illustrates relevance judgments for two queries that share a reference image and candidate set but express different retrieval objectives. We define relevance as the extent to which the candidate's visual and textual evidence satisfies the complete request under its intent.

\paragraph{Query requirements.}
The query region $b_q$ identifies the visual focus, while the image context and query text specify the requested outcome and its constraints. In Figure~\ref{fig:relevance-judgments}, ``Find the same watch model'' has Item Search intent and requires a matching model. ``Are leather straps durable?'' has Transferable Information Seeking intent and requires information about leather-strap durability. This information may concern another watch, provided that the material and usage conditions make it applicable to the request.

\paragraph{Candidate evidence.}
The document region $b_d$ identifies the candidate's visual focus, and the associated image context, title, and body provide evidence of what the note depicts or explains. In the illustrated notes, Candidate A depicts the requested watch model but describes only its appearance. Candidate B depicts a different variant and reports creasing without cracks during regular wear while keeping the leather strap dry. Candidate C provides lily-care information.

\Needspace{9\baselineskip}
\paragraph{Intent-conditioned query--document assessment.}
The candidate evidence is assessed against the query requirements jointly. Model identity is decisive for the exact-item request, whereas applicability and answer content determine whether the durability request is satisfied. Region-level matching therefore does not impose the same identity constraint across intents. We express this assessment using a shared ordinal scale with intent-specific decision criteria:
\begin{equation}
\label{eq:relevance-grades}
 y_i(q,d)\in\{0,1,2\}.
\end{equation}
\Needspace{8\baselineskip}
\textbf{Grade 0: no specific relevance.} The candidate provides no substantive correspondence to the requested target, outcome, or applicable information need.

\textbf{Grade 1: related but insufficient.} The candidate has a specific relation to the request, but fails at least one requirement necessary to satisfy it, such as the required target correspondence, requested information, or an explicit constraint.

\textbf{Grade 2: satisfies the request.} The candidate provides the requested result and meets the correspondence and constraints required by the complete query under its intent.

Accordingly, Candidates A, B, and C in Figure~\ref{fig:relevance-judgments} receive grades $(2,1,0)$ for Item Search and $(1,2,0)$ for Transferable Information Seeking. A satisfies the model requirement but lacks durability information; B fails the exact-model requirement but provides applicable wear information. C has no relevant correspondence under either request. The grade meanings remain fixed across intents, while the conditions for satisfying the request vary. Appendix~\ref{app:rubrics} specifies the criteria for each intent.

The same relevance definition connects the two retrieval stages. Representation learning uses the ordinal relations between grades to prioritize satisfying candidates, and the reranker predicts the grade from the query--candidate evidence conditioned on intent. Section~\ref{sec:embedding-ranking} defines the ranking objective, and Section~\ref{sec:embedding-eval-setup} describes evaluation using grade-2 candidates as positives.

\FloatBarrier
\Needspace{12\baselineskip}
\section{Learning Intent-Conditioned Retrieval}
\label{sec:method}
\subsection{Framework Overview}

\method combines a multimodal embedding model, a hierarchical intent recognizer, and an intent-conditioned reranker under the relevance criteria in Section~\ref{sec:relevance}. Intent-aware contrastive learning, graded ranking, and distribution distillation train the embedding model to reflect the requirements of different retrieval objectives. Query-side adaptation further refines the query representation within the fixed document space.

\begin{figure}[H]
\centering
\includegraphics[width=\linewidth]{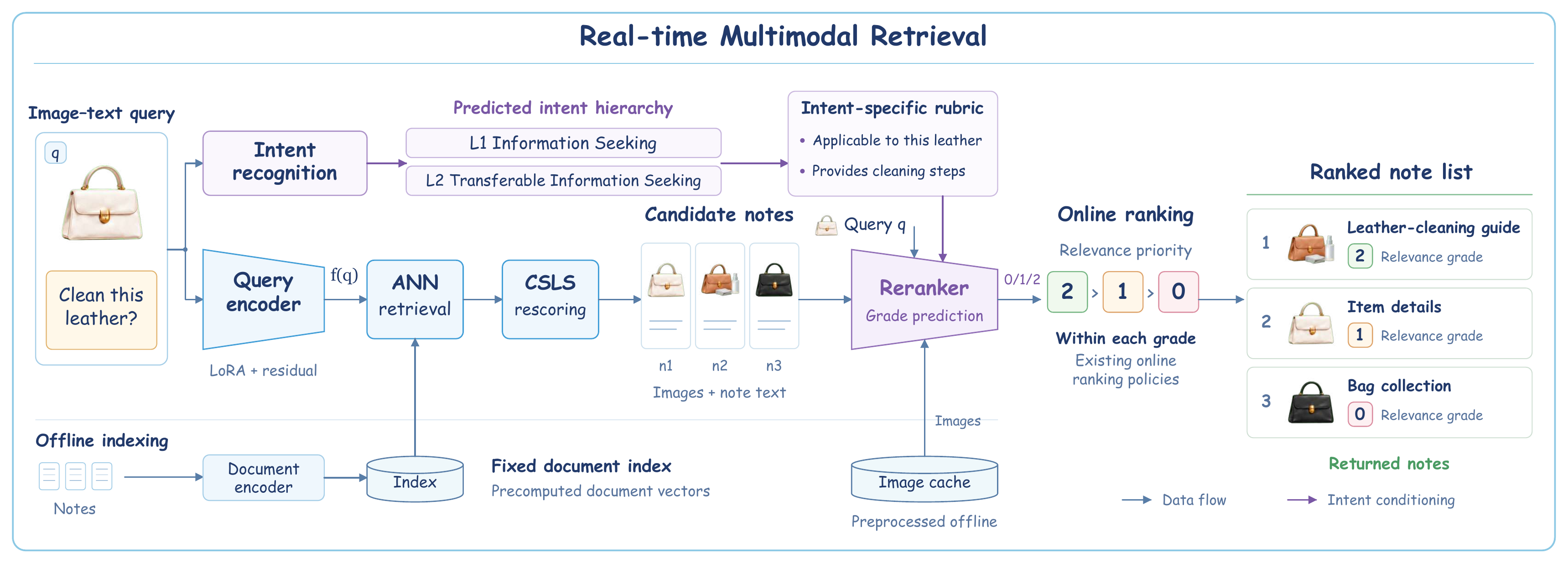}
\caption{Real-time multimodal retrieval. Query encoding and hierarchical intent recognition run concurrently. ANN retrieves candidates from the fixed document index, and CSLS adjusts their scores. The reranker uses query--candidate evidence and intent-specific criteria to predict relevance grades. Grades determine ranking priority, while existing online policies order candidates within each grade to produce the returned note list. Candidate images are prepared offline and served from a corpus-wide cache.}
\Description{A left-to-right pipeline connects the image--text query, query encoder, ANN retrieval, CSLS rescoring, candidate notes, reranker, online ranking, and returned note list. Intent recognition runs in parallel; its two-level prediction and the corresponding relevance criteria condition the reranker, which also receives the original query evidence. Offline document encoding supplies the fixed ANN index, and an image cache supplies preprocessed candidate images. The final stage prioritizes relevance grades 2, 1, and 0 and applies existing online ranking policies within each grade.}
\label{fig:framework}
\end{figure}
\nopagebreak[4]

Figure~\ref{fig:framework} illustrates the online pipeline. Given a query $q=(x_q,t_q,b_q)$, query encoding and hierarchical intent recognition run concurrently. Vector search retrieves candidates from the fixed document index, and CSLS adjusts their scores before reranking. The reranker assesses each candidate using the predicted first- and second-level intents together with the designated regions, image context, and text of the query and candidate. The predicted grades guide the final ordering of the returned notes; Section~\ref{sec:reranker} describes the serving procedure.

\Needspace{7\baselineskip}
\subsection{Hierarchical Intent Recognition}
\label{sec:intent-recognizer}
We formulate intent recognition as label-sequence generation over the hierarchy defined in Section~\ref{sec:intent-scope}. Given a query $q=(x_q,t_q,b_q)$, the recognizer $h_\psi$ predicts its first- and second-level intents from the reference image, query text, and query region:
\begin{equation}
\label{eq:intent-prediction}
 (\hat i_q^{(1)},\hat i_q^{(2)})=h_\psi(q).
\end{equation}

\paragraph{Training supervision.}
A multimodal teacher provides initial intent annotations, which are filtered and corrected to form training targets. A compact task instruction specifies the hierarchy, and the target sequence contains the first-level intent followed by the second-level intent.

\paragraph{Student training.}
We train a 0.8B multimodal generative student through full-parameter supervised fine-tuning on these targets. The intent student minimizes the negative log-likelihood of the target label sequence conditioned on the query and task instruction. The target response contains only the two intent labels. Appendix~\ref{app:intent-details} gives the response-token training objective.

\paragraph{Inference.}
The recognizer operates on the query independently of candidate retrieval. Its predicted intent pair $(\hat i_q^{(1)},\hat i_q^{(2)})$ conditions the reranker in Section~\ref{sec:reranker} when assessing query--candidate relevance.

\Needspace{5\baselineskip}
\subsection{Agentic Data Construction}
\label{sec:data-construction}

\begin{figure}[H]
\centering
\includegraphics[width=\linewidth,trim=44bp 114bp 42bp 119bp,clip]{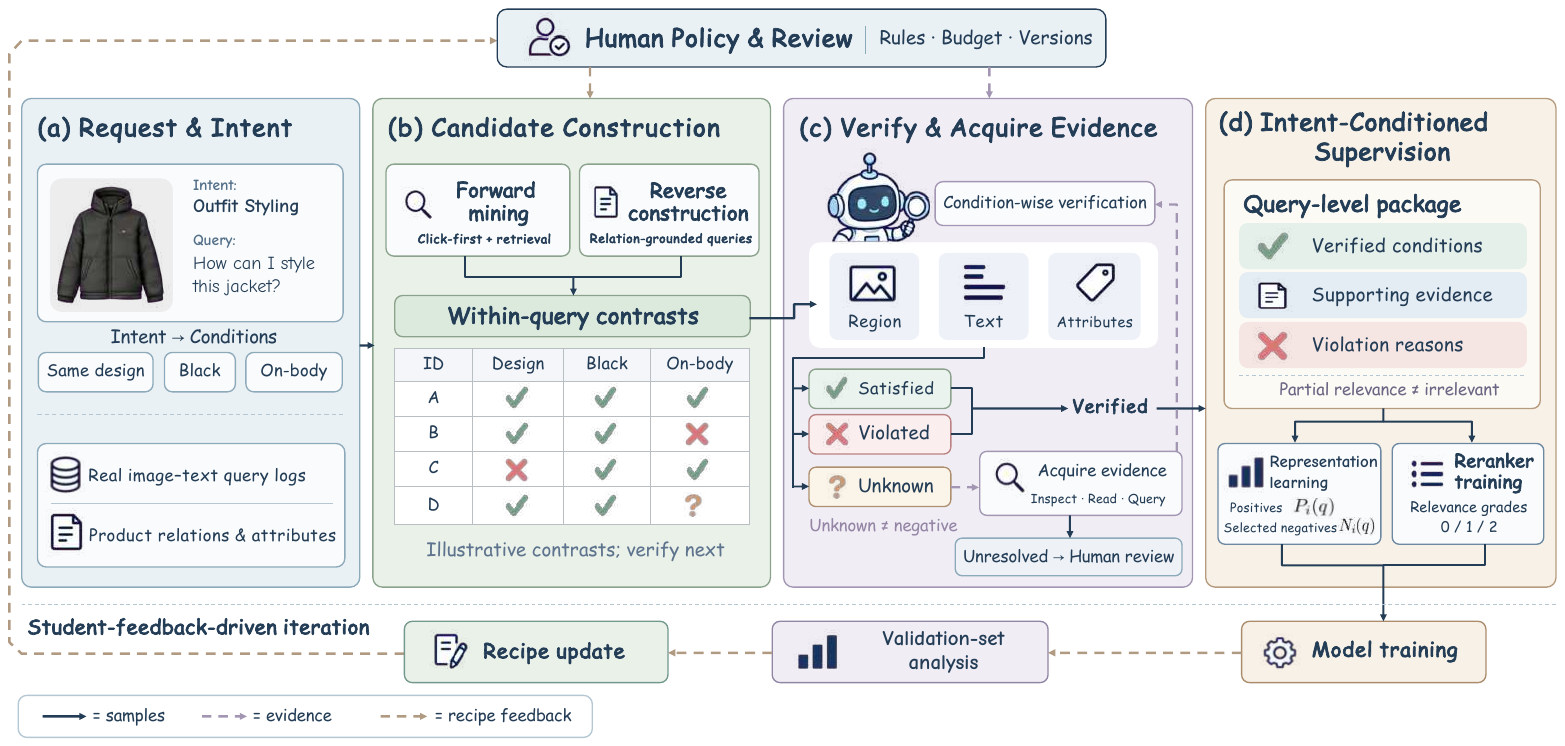}
\caption{Agentic data construction. The query and its verified intent guide forward and reverse candidate construction, with contrasts that isolate decisive requirements. Agents verify candidate evidence and acquire additional evidence for unresolved conditions. Verified examples provide supervision for representation learning and reranker training. Validation-set feedback guides targeted construction for underperforming intents and recurring errors.}
\Description{Intent-conditioned candidate construction followed by evidence verification and supervision assembly. Forward and reverse construction produce candidate contrasts. Unknown conditions trigger evidence acquisition and reassessment; ambiguous cases receive human review. Verified examples support representation learning and reranker training. Validation-set errors guide targeted construction for underperforming intents.}
\label{fig:data-construction}
\end{figure}
\nopagebreak[4]

To construct intent-consistent supervision at scale, we use an offline agentic workflow guided by the relevance criteria in Section~\ref{sec:relevance-criteria} (Figure~\ref{fig:data-construction}).

The trained intent recognizer provides initial intent annotations for source queries. The agent verifies these annotations against the query image, designated region, and textual request, and revises assignments that do not reflect the requested retrieval objective. Subsequent candidate construction and relevance assessment use the verified intents to produce supervision for representation learning and reranker training.

\paragraph{Candidate construction.}
The agents select candidate sources according to the requested outcome and constraints. Forward construction starts from real queries and combines candidates from engagement logs with multi-channel retrieval. Reverse construction supplements intents with limited candidate coverage by generating queries grounded in product attributes or relationships between notes. Both paths organize candidates around a shared query and seek contrasts on conditions that determine relevance. For the jacket example in Figure~\ref{fig:data-construction}, candidates may show the requested jacket without an outfit, or show an outfit with a different jacket. Such contrasts distinguish satisfaction of the complete request from similarity along only some dimensions.

\paragraph{Evidence verification.}
For each candidate, the agent checks the intent-specific requirements against the query and candidate regions, image context, full-note text, and available attributes. Each condition is recorded as \textbf{satisfied, violated, or unknown}. Missing evidence determines the next action, such as regional inspection, full-note reading, or attribute lookup. The agent then reassesses the unresolved conditions using the acquired evidence. Ambiguous cases receive human review. Relevance grades follow the verified evidence and the criteria in Section~\ref{sec:relevance-criteria}; judgments that remain unresolved are withheld from training supervision.

\paragraph{Training supervision.}
Verified examples associate each query--candidate pair with its intent-conditioned relevance grade, supporting evidence, and source-note identity. Grouping candidates by query yields satisfying positives $P_i(q)$, selected hard negatives $N_i(q)$, and graded candidate sets for representation learning. These labels support graded ranking, while an offline embedding teacher supplies candidate-score distributions for distillation (Section~\ref{sec:embedding-ranking}). Reranker annotation follows the same relevance criteria through the teacher-based procedure in Section~\ref{sec:reranker}. Multiple valid positives are retained, and source queries and their generated variants share a data partition.

\paragraph{Student feedback.}
We analyze student performance on the validation set by intent to identify underperforming intents and recurring errors. These findings guide targeted construction of additional examples for the affected intents and query conditions. New examples undergo the same evidence verification before inclusion in subsequent training.

\FloatBarrier

\Needspace{7\baselineskip}
\subsection{Multimodal Representation Learning}
\label{sec:embedding}

\begin{figure}[H]
\centering
\includegraphics[width=\linewidth]{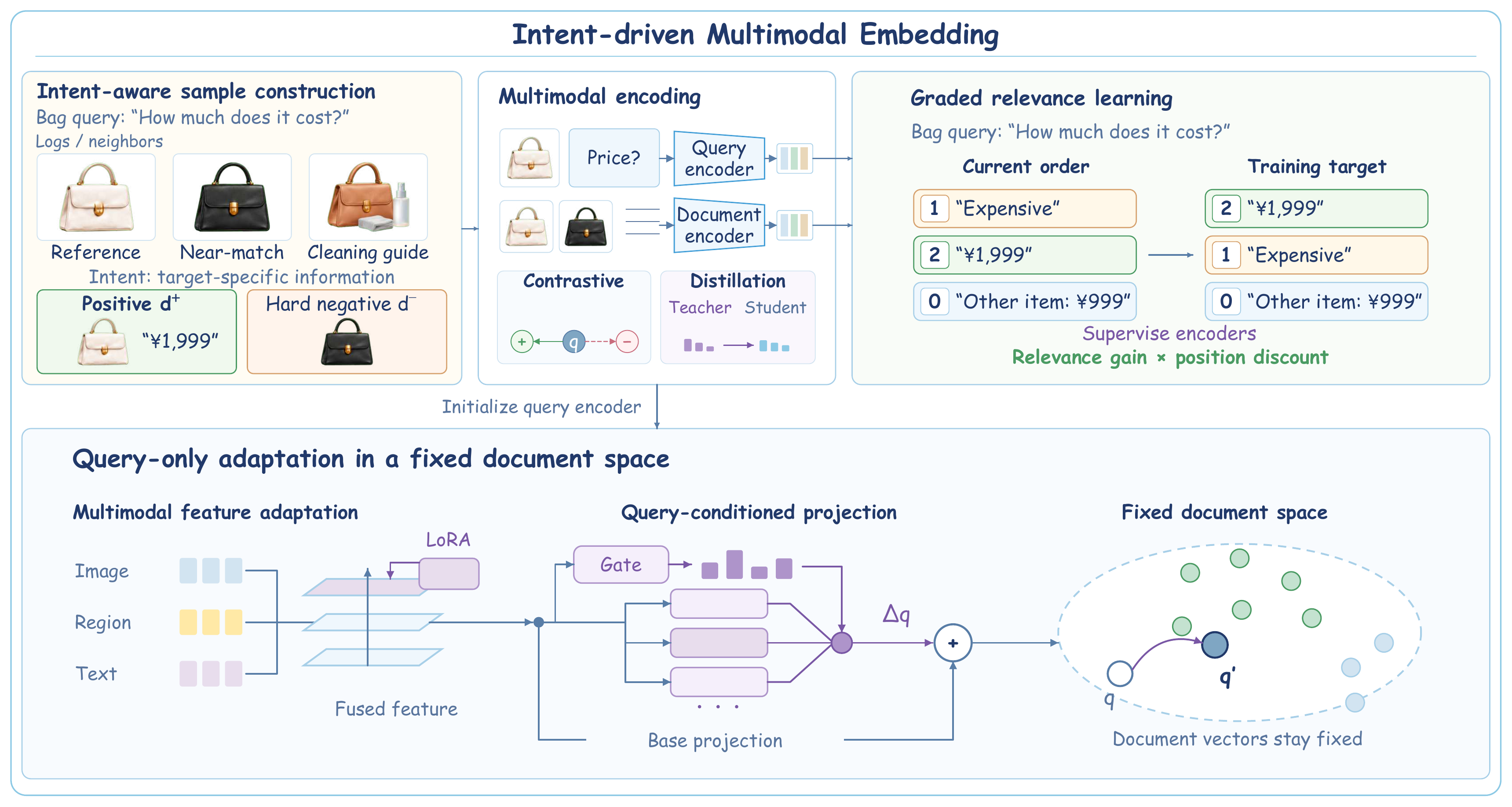}
\caption{Multimodal embedding training. Intent-specific criteria define supervision for contrastive learning, graded ranking, and candidate-distribution distillation. Query-side feature adaptation and residual projection subsequently refine retrieval within the fixed document space. The price example illustrates three-grade relevance supervision.}
\Description{The upper row shows intent-aware sample construction, query and document encoding, and graded relevance learning. Contrastive learning and teacher distribution distillation supervise the encoders. The price example ranks a specific price answer above a qualitative description and a price for another item. The lower row shows query-side LoRA feature adaptation, query-conditioned residual projection, and the base projection combined to update the query vector while keeping document vectors fixed.}
\label{fig:embedding-training}
\end{figure}
\nopagebreak[4]

Figure~\ref{fig:embedding-training} summarizes representation learning and query adaptation with a fixed document index.

The embedding model learns which query--document relationships satisfy the request and how candidates should be ordered by the extent of that satisfaction. Training proceeds from intent-dependent contrastive learning to graded ranking and knowledge distillation. After the document index is built, query-side adaptation refines multimodal features and their projection into the fixed document space.

\subsubsection{Multimodal Representation}
\label{sec:embedding-representation}
The query encoder processes $q=(x_q,t_q,b_q)$, and the document encoder processes $d=(x_d,t_d,b_d)$ as defined in Section~\ref{sec:intent-scope}. Each representation combines the designated region with its image context and associated text. Indexed representations remain associated with their source notes. A shared multimodal backbone produces a fused representation at the last valid token; separate projection heads map query and document features to the retrieval space:
\begin{equation}
\label{eq:embedding-representations}
 f_\theta(q)=\operatorname{norm}\!\left(W_qH_\theta(q)\right),
 \qquad
 g_\theta(d)=\operatorname{norm}\!\left(W_dH_\theta(d)\right).
\end{equation}
Here, $\theta$ denotes the backbone parameters, $H_\theta(\cdot)\in\mathbb R^{d_h}$ is the fused feature, and $W_q,W_d\in\mathbb R^{d_e\times d_h}$ are the query and document projections. The operator $\operatorname{norm}(u)=u/\lVert u\rVert_2$ denotes L2 normalization. The normalized embeddings $f_\theta(q),g_\theta(d)\in\mathbb R^{d_e}$ have matching score $s_\theta(q,d)=f_\theta(q)^\top g_\theta(d)$. We use $d_h=1024$ and $d_e=128$.

\subsubsection{Intent-Aware Contrastive Learning}
\label{sec:embedding-contrastive}
\textbf{Positive and hard-negative relationships} follow the criteria in Section~\ref{sec:relevance-criteria}. A positive supplies the evidence needed for the requested result and receives grade 2. A hard negative retains substantial visual or semantic similarity but fails a decisive requirement. For a price query, a note depicting the correct product without pricing information provides such a contrast; for Outfit Styling, the corresponding contrast is a matching garment without the requested combination. These examples train the representation to distinguish request satisfaction from appearance alone.

\paragraph{Contrastive objective.}
Consider a batch $\mathcal B=\{(q_a,d_a^+)\}_{a=1}^{B}$ of $B$ positive query--document pairs, where $a$ indexes a pair and $d_a^+$ is the document paired with $q_a$. Let $\mathcal C_a^D$ and $\mathcal C_a^Q$ be the valid document and query candidate sets for the two retrieval directions. These sets retain the paired positive and mask invalid identity collisions.

\Needspace{12\baselineskip}
With $n_a$ denoting an additional hard negative for $q_a$, the directional contrastive losses are
\begin{align}
 \ell_a^{q\rightarrow d}
 &=-\log\frac{\exp\!\left(s_\theta(q_a,d_a^+)/\tau_1\right)}
 {\displaystyle\sum_{d\in\mathcal C_a^D}\exp\!\left(s_\theta(q_a,d)/\tau_1\right)
  +v_aw(t)\exp\!\left(s_\theta(q_a,n_a)/\tau_1\right)},
 \label{eq:contrastive-q2d}\\
 \ell_a^{d\rightarrow q}
 &=-\log\frac{\exp\!\left(s_\theta(q_a,d_a^+)/\tau_1\right)}
 {\displaystyle\sum_{q\in\mathcal C_a^Q}\exp\!\left(s_\theta(q,d_a^+)/\tau_1\right)}.
 \label{eq:contrastive-d2q}
\end{align}
The temperature $\tau_1$ is learnable and initialized to $0.07$. The indicator $v_a\in\{0,1\}$ specifies whether the additional negative is valid. Its coefficient $w(t)$ increases gradually with training step $t$, allowing pair alignment to develop before strengthening this explicit negative term. This coefficient applies only to the additional negative, rather than to all candidates. Appendix~\ref{app:embedding-details} gives the schedule.

The bidirectional contrastive objective averages over paired examples and retrieval directions:
\begin{equation}
\label{eq:contrastive}
 \mathcal L_{\mathrm{CL}}^{(1)}
 =\frac{1}{2B}\sum_{a=1}^{B}
 \left(\ell_a^{q\rightarrow d}+\ell_a^{d\rightarrow q}\right).
\end{equation}

\paragraph{Hard-negative margin.}
Contrastive learning establishes the overall matching structure; a margin constraint further separates positives from their closest competitors. For each query, the $K_{\mathrm{hard}}$ highest-scoring valid negatives from an expanded candidate set form $\mathcal H_a^D$. The reverse-direction set $\mathcal H_a^Q$ contains high-scoring negative queries for $d_a^+$. We optimize
\begin{align}
 \ell_{\mathrm m,a}^{q\rightarrow d}
 &=\frac{1}{|\mathcal H_a^D|}
   \sum_{n\in\mathcal H_a^D}
   \left[m+s_\theta(q_a,n)-s_\theta(q_a,d_a^+)\right]_+,
 \label{eq:embedding-margin-q2d}\\
 \ell_{\mathrm m,a}^{d\rightarrow q}
 &=\frac{1}{|\mathcal H_a^Q|}
   \sum_{q^-\in\mathcal H_a^Q}
   \left[m+s_\theta(q^-,d_a^+)-s_\theta(q_a,d_a^+)\right]_+.
 \label{eq:embedding-margin-d2q}
\end{align}
Here, $|\mathcal H|$ is the number of selected negatives, $[z]_+=\max(z,0)$, and $m$ is the target score margin. We set $K_{\mathrm{hard}}=30$ and $m=0.05$. Averaging over examples and directions gives the margin objective:
\begin{equation}
\label{eq:embedding-margin}
 \mathcal L_{\mathrm{margin}}
 =\frac{1}{2B}\sum_{a=1}^{B}
   \left(\ell_{\mathrm m,a}^{q\rightarrow d}
        +\ell_{\mathrm m,a}^{d\rightarrow q}\right).
\end{equation}
The first-stage objective is
\begin{equation}
\label{eq:embedding-stage1}
 \mathcal L_{\mathrm{P1}}(t)
 =\mathcal L_{\mathrm{CL}}^{(1)}(t)
  +\lambda_m(t)\mathcal L_{\mathrm{margin}},
\end{equation}
where $\lambda_m(t)$ increases to $0.25$. Averaging the constraint over multiple hard negatives encourages separation from several confusable candidates, rather than only the single highest-scoring negative.

\subsubsection{Learning to Rank with Graded Relevance}
\label{sec:embedding-ranking}
Binary supervision separates matches from non-matches, while \textbf{graded supervision} determines which useful candidate should appear first. For a query asking the price of a bag, a note specifying CNY 1,999 fully answers the question; a note calling the same bag expensive provides only a qualitative indication. Both concern the requested product, but the former should rank higher.

We use the shared three-grade labels $y_i(q,d)$ defined in Section~\ref{sec:relevance-criteria}. For each query, confusable candidates from adjacent observed relevance levels form a comparison set $\mathcal A_q$. A pair $(a,b)\in\mathcal A_q$ means that $d_a$ should precede $d_b$, with $r_a>r_b$, where $r_a=y_i(q,d_a)$ and $r_b=y_i(q,d_b)$.

\paragraph{Pairwise ranking.}
We weight pairwise errors by their effect on normalized discounted cumulative gain (NDCG)~\citep{jarvelin2002cumulated,burges2010ranknet}. Relevance gain and position discount are
\begin{equation}
\label{eq:embedding-gain}
 \mathcal G(r)=2^r-1,\qquad
 D_L(p)=\frac{\mathbf{1}[p\le L]}{\log_2(1+p)}.
\end{equation}
Here, $r\in\{0,1,2\}$ is the relevance grade, $p$ is a one-indexed rank, $L$ is the ranking cutoff, and $\mathbf{1}[\cdot]$ is the indicator function. For candidates currently at positions $p_a$ and $p_b$, swapping them changes NDCG by
\begin{equation}
\label{eq:embedding-pair-weight}
 \Delta_{q,a,b}
 =\frac{\left|
  \big(\mathcal G(r_a)-\mathcal G(r_b)\big)
  \big(D_L(p_a)-D_L(p_b)\big)\right|}
 {\operatorname{IDCG}_{L}(q)}.
\end{equation}
The denominator $\operatorname{IDCG}_{L}(q)$ is the ideal discounted cumulative gain obtained by sorting the query's candidate set by the shared relevance grades, with gains $0$, $1$, and $3$. Pair weights and candidate positions are computed from the current ranking and held fixed during the corresponding gradient update. The ranking loss is
\begin{equation}
\label{eq:embedding-rank}
 \mathcal L_{\mathrm{rank}}
 =\frac{1}{|\mathcal Q|}
   \sum_{q\in\mathcal Q}\frac{1}{|\mathcal A_q|}
   \sum_{(a,b)\in\mathcal A_q}\Delta_{q,a,b}
   \log\!\left[1+\exp\!\left(
   \frac{s_\theta(q,d_b)-s_\theta(q,d_a)}{\tau_r}
   \right)\right].
\end{equation}
The set $\mathcal Q$ contains batch queries with valid comparisons; $|\mathcal Q|$ and $|\mathcal A_q|$ denote the numbers of such queries and selected pairs, respectively. We set the ranking temperature to $\tau_r=0.07$.

Averaging by pair count controls the contribution of queries with different numbers of comparisons while preserving the scale of $\Delta_{q,a,b}$ across queries. The exponential gain emphasizes the satisfaction boundary: grades 2 and 1 have a gain difference of 2, versus 1 for grades 1 and 0. With equal normalization, position-discount differences, and score gaps, the former incurs twice the weighted loss. Position discounting additionally prioritizes comparisons affecting the top-ranked results.

\paragraph{Knowledge distillation.}
Graded ranking specifies the order between the three relevance grades. A teacher distribution adds relative preferences across the candidate set, including distinctions between \textbf{candidates assigned the same grade}. We transfer this information through knowledge distillation~\citep{hinton2015distilling}.

Let $T$ and $S$ denote the frozen teacher and trainable student. For $M\in\{T,S\}$, define its query-to-document distribution over the valid document candidates $\mathcal C_q^D$ as
\begin{equation}
\label{eq:embedding-candidate-distribution}
 p_M(d\mid q;\tau)
 =\frac{\exp\!\left(s_M(q,d)/\tau\right)}
 {\displaystyle\sum_{d'\in\mathcal C_q^D}
  \exp\!\left(s_M(q,d')/\tau\right)}.
\end{equation}
Here, $s_M$ is model $M$'s matching score and $\tau$ controls distribution smoothness. The reverse distribution $p_M(q\mid d;\tau)$ is defined over valid query candidates $\mathcal C_d^Q$. Teacher and student use the same candidate sets. We retain bidirectional paired contrastive learning:
\begin{equation}
\label{eq:embedding-stage2-contrastive}
 \mathcal L_{\mathrm{CL}}^{(2)}
 =-\frac{1}{2B}\sum_{a=1}^{B}
 \left[\log p_S(d_a^+\mid q_a;\tau_c)
      +\log p_S(q_a\mid d_a^+;\tau_c)\right],
\end{equation}
where $\tau_c=0.07$. The distribution distillation objective is
\begin{equation}
\label{eq:embedding-distillation}
\begin{split}
 \mathcal L_{\mathrm{KD}}
 &=\frac12\,\mathbb E_q\!\left[
 \operatorname{KL}\!\left(
 p_T(\cdot\mid q;\tau_k)\,\middle\|\,p_S(\cdot\mid q;\tau_k)\right)\right]\\
 &\quad+\frac12\,\mathbb E_d\!\left[
 \operatorname{KL}\!\left(
 p_T(\cdot\mid d;\tau_k)\,\middle\|\,p_S(\cdot\mid d;\tau_k)\right)\right].
\end{split}
\end{equation}
The operator $\operatorname{KL}$ denotes Kullback--Leibler divergence. Expectations $\mathbb E_q$ and $\mathbb E_d$ average over query and document anchors; the distillation temperature is $\tau_k=0.10$. A 2B multimodal teacher supplies offline candidate distributions to the 0.8B retrieval student, keeping teacher computation outside the serving path.

The second-stage objective is
\begin{equation}
\label{eq:embedding-stage2}
 \mathcal L_{\mathrm{P2}}
 =\mathcal L_{\mathrm{CL}}^{(2)}
  +\lambda_{\mathrm{rank}}\mathcal L_{\mathrm{rank}}
  +\lambda_{\mathrm{KD}}\mathcal L_{\mathrm{KD}},
\end{equation}
where $\lambda_{\mathrm{rank}}=1.0$ and $\lambda_{\mathrm{KD}}=0.25$ weight ranking and distillation. The three terms supervise paired relevance, relevance-level ordering, and soft candidate preferences. This stage updates the shared backbone and both projection heads, producing base model $A$. Its document embeddings form the index used in the following adaptation stage.

\subsubsection{Query Adaptation with a Fixed Document Index}
\label{sec:embedding-query-adaptation}
Updating document representations requires recomputing their embeddings. We instead adapt the query encoder within the established document space. The two adaptation components act at different levels: \textbf{high-layer LoRA} adjusts which information from the full image, reference region, and text is preserved in the fused feature, while \textbf{residual projection} combines that information into an appropriate retrieval direction.

\paragraph{Query-conditioned residual projection.}
Let
\begin{equation}
\label{eq:embedding-adapted-feature}
 h_q=H_{\theta_A,\eta_{\mathrm L}}(q),\qquad u_q=W_qh_q,
\end{equation}
where $\theta_A$ denotes the frozen backbone parameters of base model $A$, $\eta_{\mathrm L}$ denotes query-side LoRA parameters, $h_q$ is the adapted feature, and $u_q$ is its base projection. A query-dependent gate mixes $K_I=6$ experts:
\begin{equation}
\label{eq:embedding-gating}
 \alpha(q)=\operatorname{softmax}\!\left(W_{\mathrm{gate}}h_q\right).
\end{equation}
Here, $W_{\mathrm{gate}}\in\mathbb R^{K_I\times d_h}$ is the gating matrix and $\alpha_k(q)$ is the weight of expert $k$. The gate is computed inside the query encoder from the fused feature. The independent hierarchical recognizer in Section~\ref{sec:intent-recognizer} continues to supply explicit labels to the reranker.

Two bottleneck branches produce residual corrections:
\begin{equation}
\label{eq:embedding-residual}
 \mathcal R_j(h_q)
 =\sum_{k=1}^{K_I}\alpha_k(q)\,U_{j,k}V_jh_q,
 \qquad j\in\{1,2\}.
\end{equation}
The branch index is $j$, $V_j\in\mathbb R^{r_j\times d_h}$ is a bottleneck projection, and $U_{j,k}\in\mathbb R^{d_e\times r_j}$ is expert $k$'s output projection. We use bottleneck ranks $r_1=8$ and $r_2=64$. The final query embedding is
\begin{equation}
\label{eq:embedding-adapted-query}
 f_\eta(q)=\operatorname{norm}\!\left[
 u_q+\beta\lVert u_q\rVert_2
 \sum_{j=1}^{2}\mathcal R_j(h_q)\right],
\end{equation}
where $\beta=0.1$ scales the residual relative to the base projection and $\eta$ collects the trainable query-adaptation parameters. The base projection is retained, while the residual adjusts the query direction according to its fused content.

\paragraph{Multimodal feature adaptation.}
We apply LoRA~\citep{hu2022lora} to the attention output and feed-forward down-projection matrices in the last $L_a=12$ fusion layers:
\begin{equation}
\label{eq:embedding-lora}
 W_\ell^{\mathrm{query}}
 =W_\ell+\frac{\gamma_{\mathrm L}}{r_{\mathrm L}}B_\ell A_\ell.
\end{equation}
For an adapted matrix $W_\ell\in\mathbb R^{d_{\mathrm{out}}\times d_{\mathrm{in}}}$, $\ell$ indexes its layer, $A_\ell\in\mathbb R^{r_{\mathrm L}\times d_{\mathrm{in}}}$ and $B_\ell\in\mathbb R^{d_{\mathrm{out}}\times r_{\mathrm L}}$ are trainable factors, and the original weight $W_\ell$ remains frozen. We use LoRA rank $r_{\mathrm L}=8$ and scaling coefficient $\gamma_{\mathrm L}=16$.

\paragraph{Joint optimization.}
We jointly optimize $\eta=\{A_\ell,B_\ell,U_{j,k},V_j,W_{\mathrm{gate}}\}$ with the contrastive, ranking, and distillation objectives in Equation~\eqref{eq:embedding-stage2}, replacing the student score with
\begin{equation}
\label{eq:embedding-fixed-index-score}
 s_\eta(q,d)=f_\eta(q)^\top g_A(d).
\end{equation}
Here, $g_A(d)$ is the fixed document embedding from base model $A$. Ranking supervision backpropagates through the residual projection into query-side LoRA, refining both the retained multimodal information and its mapping to the index.

\paragraph{Online inference.}
The document path disables query-specific adapters and retains its existing embeddings. Serving requires one query-encoding pass and a \textbf{single 128-dimensional vector}.

\paragraph{Candidate score correction.}
We apply cross-domain similarity local scaling (CSLS)~\citep{conneau2018word} to the initial top-$K_c$ candidates. For a fixed query, the query-dependent density term is constant across candidates, so the corrected ranking can use
\begin{equation}
\label{eq:embedding-csls}
 \widetilde s(q,d)=s_\eta(q,d)
 -\frac{1}{2k}\sum_{q'\in\mathcal N_Q^k(d)}s_\eta(q',d).
\end{equation}
The set $\mathcal N_Q^k(d)$ contains document $d$'s $k$ nearest queries from a training-query reference bank. We use $k=10$ and $K_c=500$. The document-specific correction is computed offline for the adapted query model, leaving only score adjustment during serving. The corrected candidates proceed to the intent-conditioned reranker.

\FloatBarrier
\Needspace{7\baselineskip}
\subsection{Intent-Conditioned Reranker}
\label{sec:reranker}

We formulate reranking as intent-conditioned relevance prediction for a query $q=(x_q,t_q,b_q)$ and a candidate document unit $d=(x_d,t_d,b_d)$, following the definitions in Section~\ref{sec:intent-scope}. The reranker jointly assesses the regions specified by $b_q$ and $b_d$, their image context, and the associated text $t_q$ and $t_d$, where $t_d$ contains the candidate note's title and body. Conditioned on the query's first- and second-level intents, $i_q^{(1)}$ and $i_q^{(2)}$, it predicts the relevance grade $y_i(q,d)\in\{0,1,2\}$ in Section~\ref{sec:relevance-criteria}, with $i=i_q^{(2)}$. The prompt includes the relevance criteria for the supplied intent.

\paragraph{Teacher annotation.}
Teacher annotations apply the assessment in Section~\ref{sec:relevance-criteria}, evaluating candidate visual and textual evidence against the requirements expressed by the complete query under its intent. Two teacher models independently assign a relevance grade and an evidence-based explanation to each query--candidate pair. When their grades agree, the shared grade is adopted; otherwise, a third teacher adjudicates the pair to determine its final relevance label. The resulting labels supervise student training.

\begin{figure}[H]
\centering
\includegraphics[width=\linewidth]{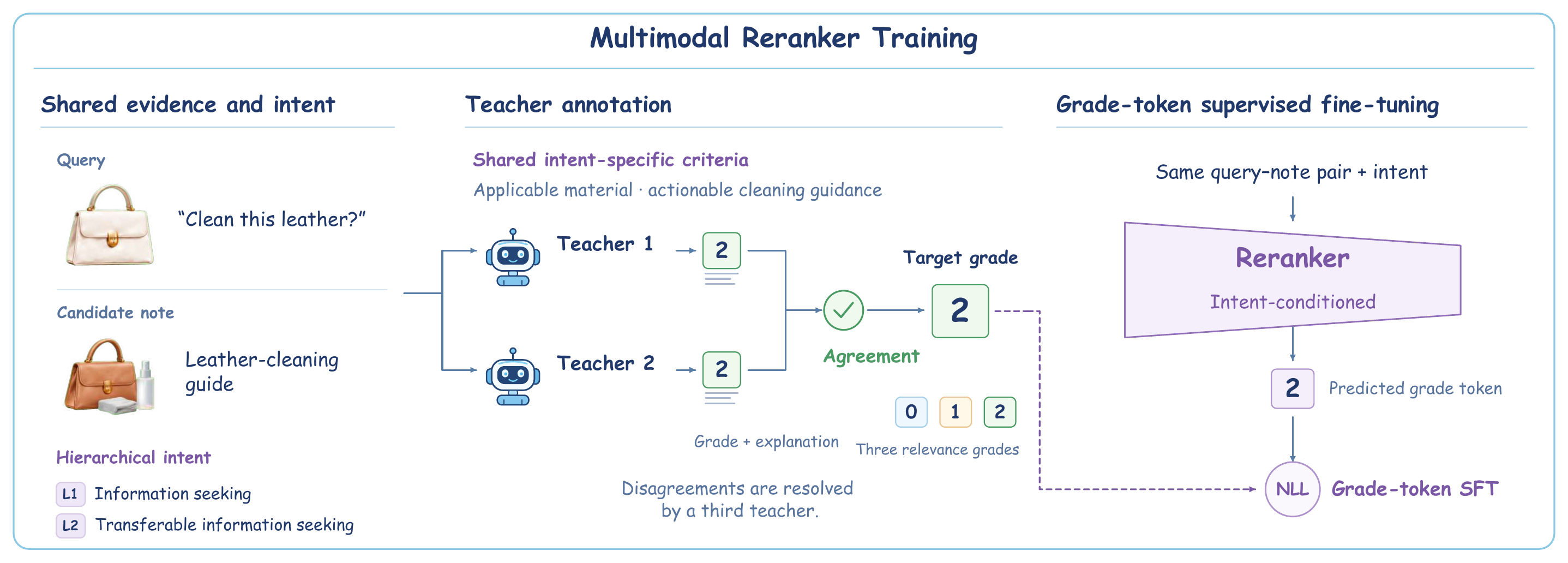}
\caption{Multimodal reranker training. Two teachers independently assess shared query--candidate evidence under intent-specific relevance criteria. Agreement or third-teacher adjudication determines the target grade. The student receives the query--candidate evidence and intent and is trained to predict the grade token by supervised fine-tuning.}
\Description{Shared query--candidate evidence and hierarchical intent support two independent teacher assessments under the same intent-specific criteria. Teacher agreement or third-teacher adjudication determines the target grade in $\{0,1,2\}$. The student reranker receives the same evidence and intent and minimizes the negative log-likelihood of the target grade token. Teacher explanations support annotation but are not student training targets.}
\label{fig:reranker-training}
\end{figure}
\nopagebreak[4]

Figure~\ref{fig:reranker-training} summarizes the teacher-supervised training workflow.

\paragraph{Student training.}
We train a 0.8B multimodal generative student through full-parameter supervised fine-tuning using the relevance labels obtained through teacher agreement and third-teacher adjudication. Conditioned on the query--candidate pair and the intent labels provided for the query, the relevance student minimizes the negative log-likelihood of the target relevance label. The target response contains \textbf{only the relevance label}; teacher explanations are not included. Appendix~\ref{app:reranker-objective} gives the training objective.

\paragraph{Real-time inference.}
At inference, the reranker receives the first- and second-level intents predicted by the recognizer in Section~\ref{sec:intent-recognizer}, $(\hat i_q^{(1)},\hat i_q^{(2)})$, and evaluates the original query--candidate image--text evidence to assign each candidate a relevance grade in $\{0,1,2\}$. Candidates are ordered by decreasing relevance grade, prioritizing results that satisfy the request over related but insufficient results and irrelevant results. Existing online ranking policies determine the order within each grade, producing the returned note list.

To support real-time serving, we maintain a corpus-wide image cache through historical backfill and continuous incremental updates. Images are resized and converted to a standard format offline, then stored in low-latency object storage. The serving system retrieves cached images by identifier for model input preparation. On a cache miss, it fetches the image from the source, processes it, and populates the cache. Reusing processed images across requests reduces repeated preprocessing and origin access, helping control serving overhead and tail latency. Together with precomputed document embeddings and compact online models, this infrastructure supports real-time multimodal search at industrial scale.

\FloatBarrier
\section{Experimental Evaluation}

\label{sec:experiments}

We evaluate \method from three perspectives: online system effectiveness, component performance, and serving efficiency. Section~\ref{sec:overall-eval} reports online results for two cumulative production configurations. Sections~\ref{sec:intent-eval}--\ref{sec:relevance-eval} examine hierarchical intent recognition, multimodal retrieval, and intent-conditioned relevance prediction. Section~\ref{sec:efficiency} analyzes the execution design and serving latency.

\subsection{Experimental Setup}

\label{sec:experimental-setup}

\paragraph{Online evaluation.}
We conduct an online A/B test on Xiaohongshu multimodal search. Traffic is partitioned at the user level. Two cumulative configurations are evaluated against the same production control, which uses image-based candidate retrieval followed by reranking. All online results are reported as relative percentage changes, computed as $100(M_{\mathrm{variant}}/M_{\mathrm{control}}-1)$, where $M$ denotes the corresponding metric.

\paragraph{Online metrics.}
Top-4 query click-through rate (Top-4 QCTR) is the fraction of issued queries with at least one clicked note among the first four results. Query click-through rate (QCTR) is the fraction with at least one clicked result at any rank. Average time spent on search results pages per user measures the cumulative time spent on search results pages during the evaluation period, averaged across users. These metrics characterize click incidence at the top ranks, click incidence across the returned results, and user engagement time, respectively.

\paragraph{Component metrics.} Intent recognition is evaluated by first- and second-level accuracy and second-level macro-F1. Relevance prediction is evaluated by accuracy on three-grade relevance classification (0 / 1 / 2), binary satisfaction classification (2 versus 0--1), and binary relevance classification (1--2 versus 0), all derived from the same predictions. Multimodal retrieval uses Hit@1, Hit@10, Hit@100, Recall@100, and NDCG@10, with fully satisfying candidates as positives (Section~\ref{sec:embedding-eval-setup}). Serving measurements report P50 and P90 latency.

In tables comparing models or configurations, gray shading identifies the model or configuration adopted in our method, and bold metric values indicate the best result within each comparison group.

\subsection{Overall Online Effectiveness}

\label{sec:overall-eval}

We evaluate two cumulative configurations of the components described in Section~\ref{sec:method}. The first combines hierarchical intent recognition with intent-conditioned reranking; these modules are evaluated jointly because the reranker receives the recognizer's intent predictions. The second additionally incorporates multimodal retrieval to form the deployed \method configuration. Table~\ref{tab:overall} reports both configurations relative to the same production control.

\begin{table}[htbp]
\centering\reporttablefont
\caption{Online results from a user-level A/B test. All values are relative percentage changes against the same production control. The second configuration adds multimodal retrieval to the first.}
\label{tab:overall}
\begin{tabularx}{\textwidth}{@{}Yrrr@{}}
\toprule
\reporttablehead{Configuration} & \multicolumn{1}{c}{\reporttablehead{Top-4 QCTR}} & \multicolumn{1}{c}{\reporttablehead{QCTR}} & \shortstack[c]{\reporttablehead{Time spent on search}\\\reporttablehead{results pages per user}} \\
\midrule
Intent recognition and reranking & +21.65\% & +15.12\% & +9.66\% \\
\addlinespace
\rowcolor{black!6}
\hspace*{0.65em}\textbf{+ Multimodal retrieval (\method)} & \textbf{+28.17\%} & \textbf{+18.96\%} & \textbf{+12.56\%} \\
\bottomrule
\end{tabularx}
\end{table}

\paragraph{Online results.}
The configuration with intent recognition and reranking improves Top-4 QCTR by 21.65\%, QCTR by 15.12\%, and average time spent on search results pages per user by 9.66\% relative to the production control. Adding multimodal retrieval yields corresponding gains of 28.17\%, 18.96\%, and 12.56\%. The higher gains across all three metrics support combining multimodal candidate retrieval with intent-conditioned relevance assessment to improve online search engagement.

\subsection{Hierarchical Intent Recognition}

\label{sec:intent-eval}

\paragraph{Data and evaluation.}
The intent student is trained on 26,767 labeled queries and evaluated on 390 human-annotated queries. Training labels follow the annotation and correction procedure in Section~\ref{sec:intent-recognizer}. Accuracy is computed separately for the first- and second-level predictions, and second-level macro-F1 averages the F1 scores of the seven intents.

\begin{table}[htbp]
\centering\reporttablefont
\caption{Hierarchical intent recognition on 390 human-labeled evaluation queries. All values are percentages.}
\label{tab:intent-results}
\begin{tabularx}{\textwidth}{@{}Yrrr@{}}
\toprule
\reporttablehead{Model} & \shortstack{\reporttablehead{First-level}\\\reporttablehead{accuracy}} & \shortstack{\reporttablehead{Second-level}\\\reporttablehead{accuracy}} & \shortstack{\reporttablehead{Second-level}\\\reporttablehead{macro-F1}} \\
\midrule
Teacher & \textbf{96.15} & \textbf{91.79} & \textbf{91.84} \\
\rowcolor{black!6}
Intent student (0.8B) & 95.90 & 91.54 & 91.63 \\
\bottomrule
\end{tabularx}
\end{table}

The 0.8B intent student closely approaches the teacher in both first- and second-level accuracy and in second-level macro-F1, retaining comparable aggregate recognition performance with a compact model (Table~\ref{tab:intent-results}).

\paragraph{Error analysis.}
Fifteen of the student's 33 second-level errors involve Target-Specific Information Seeking and Transferable Information Seeking. This boundary is particularly challenging within the current intent taxonomy: both categories seek textual information, but they differ in whether the answer must concern the designated target or can apply across targets. Distinguishing them requires interpreting the scope of the information need and its dependence on the reference target, rather than recognizing a different topic or object category. The distinction also affects downstream relevance judgments, since evidence sufficient for a transferable request may not satisfy a target-specific one.

\FloatBarrier
\subsection{Multimodal Retrieval}

\label{sec:embedding-eval}

Multimodal retrieval must combine image content, the reference region, and textual constraints to identify notes that meet the requested objective. Beyond distinguishing relevant from irrelevant candidates, the embedding must learn which candidates should appear first. We evaluate these capabilities through overall retrieval results, intent-level analyses, training-objective comparisons, and query adaptation with a fixed document index.

\subsubsection{Retrieval Evaluation Setup}

\label{sec:embedding-eval-setup}

We evaluate on an independent test set of 20,000 queries against two galleries containing 512,655 and 1,904,054 notes, referred to as the 0.5M and 1.9M galleries. Both galleries use the same test queries, which are excluded from training.

Only \textbf{fully satisfying candidates}, with shared relevance grade $y_i(q,d)=2$, count as positives in retrieval evaluation. We report \textbf{Hit@1, Hit@10, Hit@100, Recall@100, and NDCG@10}. Hit@$K$ measures the fraction of queries with at least one positive note among the top-$K$ results. Recall@100 averages the per-query fraction of positive notes retrieved within the top 100, giving every query equal weight. Evaluation NDCG@10 uses binary gains for grade-2 positives to measure their discounted placement within the top ten results, normalized by the ideal ranking. The training objective in Section~\ref{sec:embedding-ranking} uses graded gains to learn the order among all three relevance levels.

\paragraph{Controlled supervision comparison.}
We compare generic and intent-aware relevance supervision while holding all other data-construction and training conditions fixed. Both variants use the same source data, automated candidate-acquisition and verification workflow, data budget, model initialization, first-stage training objective, optimization settings, and training budget. They differ only in the relevance rules used to construct supervision: \textbf{Generic} applies generic matching rules, whereas \textbf{Intent-aware} applies the intent-specific criteria in Section~\ref{sec:relevance-criteria}. These rules determine the positive and hard-negative relationships used in training. Subsequent configurations add graded ranking and distillation, fixed-index query adaptation, and CSLS score correction to the intent-aware model. Appendix~\ref{app:embedding-details} summarizes the training settings.

\paragraph{External baselines.}
We include WeMM-Embedding-2B~\citep{zhou2026wemm} as an external multimodal embedding baseline, using both its full 2,048-dimensional representation and a 128-dimensional configuration matching our retrieval-vector dimension. The lower-dimensional vectors are obtained through Matryoshka prefix truncation followed by L2 normalization. The reported evaluation uses 20,000 queries and galleries containing 512,655 and 1,904,054 notes, with exact inner-product retrieval. Appendix~\ref{app:wemm-evaluation} provides the baseline configuration and the complete dimension sweep.

\subsubsection{Overall Retrieval Performance}

\label{sec:embedding-eval-overall}

Table~\ref{tab:embedding-overall} presents the external baselines alongside our internal configurations. The external comparison evaluates retrieval performance at different model sizes and embedding dimensions. Within the internal block, Generic and Intent-aware isolate the effect of data structured by intent under the controlled setting in Section~\ref{sec:embedding-eval-setup}; subsequent rows show cumulative additions to the intent-aware model.

\begin{table}[htbp]
\centering
\fontsize{9}{11}\selectfont
\setlength{\tabcolsep}{3pt}
\renewcommand{\arraystretch}{1.22}
\caption{Overall multimodal retrieval performance with external baselines and internal configurations. Subsequent rows within the internal block cumulatively add components to the intent-aware model. All metric values are percentages.}
\label{tab:embedding-overall}
\begin{tabularx}{\textwidth}{@{}>{\raggedright\arraybackslash}Xcc*{3}{r}@{\hspace{7pt}}*{3}{r}@{}}
\toprule
\multirow{2}{=}{\textbf{Model / configuration}} & \multirow{2}{*}{\textbf{Params}} & \multirow{2}{*}{\textbf{Dim}} & \multicolumn{3}{c}{\shortstack{\textbf{0.5M candidate pool}\\(512,655 notes)}} & \multicolumn{3}{c@{}}{\shortstack{\textbf{1.9M candidate pool}\\(1,904,054 notes)}} \\
\cmidrule(lr){4-6}\cmidrule(l){7-9}
 & & & \multicolumn{1}{c}{\retrievalmetric{Hit@1}} & \multicolumn{1}{c}{\retrievalmetric{Hit@10}} & \multicolumn{1}{c}{\retrievalmetric{Hit@100}} & \multicolumn{1}{c}{\retrievalmetric{Hit@1}} & \multicolumn{1}{c}{\retrievalmetric{Hit@10}} & \multicolumn{1}{c}{\retrievalmetric{Hit@100}} \\
\midrule
\multicolumn{9}{@{}l}{\textit{External baselines}}\\[2pt]
WeMM-Embedding & 2B & 128 & 37.95 & 65.81 & 85.31 & 25.42 & 52.48 & 75.27 \\
WeMM-Embedding & 2B & 2,048 & \textbf{46.14} & \textbf{73.59} & \textbf{89.22} & \textbf{32.26} & \textbf{60.94} & \textbf{81.06} \\
\midrule
\multicolumn{9}{@{}l}{\textit{Our configurations}}\\[2pt]
Generic relevance supervision & 0.8B & 128 & 60.87 & 83.37 & 92.99 & 45.51 & 74.23 & 87.67 \\
Intent-aware relevance supervision & 0.8B & 128 & 80.60 & 96.86 & 99.55 & 66.09 & 92.07 & 98.57 \\
\hspace*{0.65em}+ Graded ranking and distillation & 0.8B & 128 & 81.62 & 97.00 & 99.56 & 67.70 & 92.43 & 98.63 \\
\hspace*{0.65em}+ Fixed-index query adaptation & 0.8B & 128 & 82.22 & 97.17 & 99.56 & 68.54 & 92.79 & 98.72 \\
\rowcolor{black!6}
\hspace*{0.65em}+ CSLS score correction & 0.8B & 128 & \textbf{82.65} & \textbf{97.37} & \textbf{99.62} & \textbf{68.84} & \textbf{93.01} & \textbf{98.85} \\
\bottomrule
\end{tabularx}
\par\vspace{3pt}
{\fontsize{7.5}{9}\selectfont\raggedright Params: model size; Dim: retrieval-vector dimension.\par}
\end{table}

\textbf{Comparison with WeMM.} Our 0.8B retriever with 128-dimensional embeddings outperforms both evaluated WeMM-Embedding-2B configurations across all reported metrics in both galleries. On the 1.9M gallery, our final configuration achieves 68.84\% Hit@1, compared with 32.26\% for WeMM at its full 2,048-dimensional output, a difference of 36.58 percentage points. On the 0.5M gallery, the corresponding scores are 82.65\% and 46.14\%. These results show the effectiveness of the multimodal representation model based on structured intent in this application setting.

\textbf{Effect of data structured by intent.} Under identical data-construction and first-stage training conditions, replacing generic relevance rules with intent-specific rules raises Hit@1 from 45.51\% to 66.09\% on the 1.9M gallery and from 60.87\% to 80.60\% on the 0.5M gallery. The gains of 20.58 and 19.73 percentage points, respectively, demonstrate the value of constructing supervision around the requested outcome rather than generic matching criteria.

\textbf{Second-stage fine-grained retrieval optimization.} Graded ranking and distillation further improve Hit@1 to 67.70\% on the 1.9M gallery. Subsequent fixed-index query adaptation increases it to 68.54\%, followed by 68.84\% after CSLS score correction. These additions primarily improve the placement of satisfying candidates near the top of the ranking, with Hit@100 already reaching 98.57\% under intent-aware supervision.

\subsubsection{Effect of Intent-Aware Supervision}

\label{sec:embedding-eval-intent}

Different intents require different evidence of relevance. Item and image search emphasize the requested identity, attribute-variant search requires a specified change, and outfit requests require the intended styling relation. Information-seeking queries additionally require an answer or applicable guidance. Table~\ref{tab:embedding-by-intent} applies the controlled supervision comparison from Section~\ref{sec:embedding-eval-setup} to each intent: the data-construction workflow and training conditions remain fixed, and only the relevance rules change.

\begin{table}[htbp]
\centering
\fontsize{8.2}{10}\selectfont
\setlength{\tabcolsep}{1.5pt}
\renewcommand{\arraystretch}{1.22}
\caption{Effect of intent-specific relevance rules across query intents. Generic and Intent-aware share the same data-construction workflow and training conditions; only the rules used to construct supervision differ. Item Search and Image Search contain same-product and same-image queries, respectively. All values are percentages.}
\label{tab:embedding-by-intent}
\begin{tabularx}{\textwidth}{@{}>{\raggedright\arraybackslash}Xc*{5}{r}@{\hspace{5pt}}*{5}{r}@{}}
\toprule
\multirow{2}{=}{\textbf{Query intent}} & \multirow{2}{*}{\textbf{Supervision}} & \multicolumn{5}{c}{\shortstack{\textbf{0.5M candidate pool}\\(512,655 notes)}} & \multicolumn{5}{c@{}}{\shortstack{\textbf{1.9M candidate pool}\\(1,904,054 notes)}} \\
\cmidrule(lr){3-7}\cmidrule(l){8-12}
 &  & \multicolumn{1}{c}{\retrievalmetric{Hit@1}} & \multicolumn{1}{c}{\retrievalmetric{Hit@10}} & \multicolumn{1}{c}{\retrievalmetric{Hit@100}} & \multicolumn{1}{c}{\retrievalmetric{Recall@100}} & \multicolumn{1}{c}{\retrievalmetric{NDCG@10}} & \multicolumn{1}{c}{\retrievalmetric{Hit@1}} & \multicolumn{1}{c}{\retrievalmetric{Hit@10}} & \multicolumn{1}{c}{\retrievalmetric{Hit@100}} & \multicolumn{1}{c}{\retrievalmetric{Recall@100}} & \multicolumn{1}{c}{\retrievalmetric{NDCG@10}} \\
\midrule
 & Generic & 25.42 & 44.78 & 71.32 & 68.67 & 31.80 & 16.25 & 32.15 & 52.57 & 48.75 & 21.21 \\
\rowcolor{black!6}
\cellcolor{white}\multirow{-2}{=}{\raggedright Item Search} & Intent-aware & \textbf{74.02} & \textbf{91.80} & \textbf{98.47} & \textbf{97.50} & \textbf{81.12} & \textbf{59.09} & \textbf{82.02} & \textbf{95.98} & \textbf{94.23} & \textbf{68.11} \\
\cmidrule(lr){1-12}
 & Generic & 74.72 & 94.86 & 99.30 & 99.30 & 85.01 & 54.32 & 83.26 & 96.16 & 96.16 & 68.70 \\
\rowcolor{black!6}
\cellcolor{white}\multirow{-2}{=}{\raggedright Image Search} & Intent-aware & \textbf{97.21} & \textbf{99.91} & \textbf{100.00} & \textbf{100.00} & \textbf{98.84} & \textbf{85.70} & \textbf{99.83} & \textbf{99.91} & \textbf{99.91} & \textbf{93.65} \\
\cmidrule(lr){1-12}
 & Generic & 9.76 & 31.71 & 78.05 & 78.05 & 20.11 & 4.88 & 19.51 & 39.02 & 39.02 & 10.63 \\
\rowcolor{black!6}
\cellcolor{white}\multirow{-2}{=}{\raggedright Attribute Variant Search} & Intent-aware & \textbf{65.85} & \textbf{90.24} & \textbf{100.00} & \textbf{100.00} & \textbf{76.99} & \textbf{29.27} & \textbf{75.61} & \textbf{95.12} & \textbf{95.12} & \textbf{52.30} \\
\cmidrule(lr){1-12}
 & Generic & 5.93 & 19.34 & 45.67 & 45.67 & 11.79 & 2.49 & 9.49 & 24.44 & 24.44 & 5.45 \\
\rowcolor{black!6}
\cellcolor{white}\multirow{-2}{=}{\raggedright Outfit Styling} & Intent-aware & \textbf{60.14} & \textbf{87.90} & \textbf{97.75} & \textbf{97.75} & \textbf{73.92} & \textbf{40.21} & \textbf{73.43} & \textbf{91.34} & \textbf{91.34} & \textbf{55.83} \\
\cmidrule(lr){1-12}
 & Generic & 61.56 & 88.54 & 97.36 & 68.83 & 53.22 & 44.65 & 78.57 & 93.78 & 52.77 & 37.82 \\
\rowcolor{black!6}
\cellcolor{white}\multirow{-2}{=}{\raggedright \shortstack[l]{Target-Specific\\Information Seeking}} & Intent-aware & \textbf{83.53} & \textbf{98.14} & \textbf{99.80} & \textbf{93.73} & \textbf{79.67} & \textbf{70.65} & \textbf{95.14} & \textbf{99.37} & \textbf{85.61} & \textbf{65.90} \\
\cmidrule(lr){1-12}
 & Generic & \textbf{80.56} & 96.02 & 98.56 & 78.63 & 65.79 & \textbf{65.30} & \textbf{90.99} & 97.25 & 66.24 & 52.41 \\
\rowcolor{black!6}
\cellcolor{white}\multirow{-2}{=}{\raggedright \shortstack[l]{Transferable\\Information Seeking}} & Intent-aware & 76.16 & \textbf{96.83} & \textbf{99.60} & \textbf{94.12} & \textbf{73.72} & 58.46 & 90.55 & \textbf{98.75} & \textbf{85.97} & \textbf{55.80} \\
\midrule
 & Generic & 60.87 & 83.37 & 92.99 & 71.96 & 54.14 & 45.51 & 74.23 & 87.67 & 56.87 & 40.04 \\
\rowcolor{black!6}
\cellcolor{white}\multirow{-2}{=}{\raggedright Overall} & Intent-aware & \textbf{80.60} & \textbf{96.86} & \textbf{99.55} & \textbf{94.73} & \textbf{79.24} & \textbf{66.09} & \textbf{92.07} & \textbf{98.57} & \textbf{87.62} & \textbf{64.84} \\
\bottomrule
\end{tabularx}
\end{table}

\textbf{Identity and query constraints.} On the 1.9M gallery, intent-aware supervision improves Hit@1 over generic supervision by 42.84 percentage points for Item Search, 24.39 points for Attribute Variant Search, and 37.72 points for Outfit Styling. These gains show that intent-specific relevance rules improve retrieval for requests governed by item identity, attribute changes, and styling relations.

\textbf{Information seeking.} For Target-Specific Information Seeking, Recall@100 increases from 52.77\% to 85.61\%, while NDCG@10 rises from 37.82\% to 65.90\% on the larger gallery. For Transferable Information Seeking, Recall@100 improves from 66.24\% to 85.97\% and NDCG@10 from 52.41\% to 55.80\%. The latter intent presents a particular challenge: useful information may concern objects of different brands, models, or appearances, provided that it answers the request and applies under the specified conditions. The model must therefore identify knowledge that transfers across instances and distinguish it from topically related evidence that does not apply. For this intent, intent-aware supervision increases Hit@100 from 97.25\% to 98.75\%, but decreases Hit@1 from 65.30\% to 58.46\%, indicating that improved coverage of useful information has not fully translated into accurate selection of the first result. To refine this ordering, subsequent representation learning uses graded ranking supervision to distinguish different levels of request satisfaction, while the reranker combines explicit intent with query--candidate visual and textual evidence to assess applicability and prioritize useful retrieved content.

\subsubsection{Ablation of Training Objectives}

\label{sec:embedding-eval-objectives}

\paragraph{First-stage discriminative learning.}

The first stage combines bidirectional contrastive learning with explicit hard negatives. The Top-$K$ hard-negative margin objective further separates each positive from multiple high-scoring negative candidates. Table~\ref{tab:embedding-stage1-ablation} evaluates this additional constraint while retaining explicit hard negatives in both configurations.

\begin{table}[htbp]
\centering
\fontsize{8.5}{10.5}\selectfont
\setlength{\tabcolsep}{2pt}
\renewcommand{\arraystretch}{1.22}
\caption{Effect of the Top-$K$ hard-negative margin objective. Both configurations use bidirectional contrastive learning and explicit hard negatives. All values are percentages.}
\label{tab:embedding-stage1-ablation}
\begin{tabularx}{\textwidth}{@{}>{\raggedright\arraybackslash}X*{5}{r}@{\hspace{5pt}}*{5}{r}@{}}
\toprule
\multirow{2}{=}{\textbf{Top-$K$ margin}} & \multicolumn{5}{c}{\shortstack{\textbf{0.5M candidate pool}\\(512,655 notes)}} & \multicolumn{5}{c@{}}{\shortstack{\textbf{1.9M candidate pool}\\(1,904,054 notes)}} \\
\cmidrule(lr){2-6}\cmidrule(l){7-11}
 & \multicolumn{1}{c}{\retrievalmetric{Hit@1}} & \multicolumn{1}{c}{\retrievalmetric{Hit@10}} & \multicolumn{1}{c}{\retrievalmetric{Hit@100}} & \multicolumn{1}{c}{\retrievalmetric{Recall@100}} & \multicolumn{1}{c}{\retrievalmetric{NDCG@10}} & \multicolumn{1}{c}{\retrievalmetric{Hit@1}} & \multicolumn{1}{c}{\retrievalmetric{Hit@10}} & \multicolumn{1}{c}{\retrievalmetric{Hit@100}} & \multicolumn{1}{c}{\retrievalmetric{Recall@100}} & \multicolumn{1}{c}{\retrievalmetric{NDCG@10}} \\
\midrule
Without margin & 77.21 & 96.10 & 99.46 & 91.96 & 75.36 & 61.40 & 90.40 & 98.21 & 83.25 & 59.94 \\
\rowcolor{black!6}
With margin & \textbf{80.60} & \textbf{96.86} & \textbf{99.55} & \textbf{94.73} & \textbf{79.24} & \textbf{66.09} & \textbf{92.07} & \textbf{98.57} & \textbf{87.62} & \textbf{64.84} \\
\bottomrule
\end{tabularx}
\end{table}

The margin objective improves Hit@1 from 77.21\% to 80.60\% on the 0.5M gallery and from 61.40\% to 66.09\% on the 1.9M gallery. On the larger gallery, NDCG@10 increases from 59.94\% to 64.84\%, accompanied by an increase in Recall@100 from 83.25\% to 87.62\%. These results show the benefit of explicitly separating positives from their strongest negative competitors.

\paragraph{Second-stage graded ranking and distillation.}

The second stage learns preferences among candidates with different degrees of request satisfaction. For a query asking the price of an item, a note stating its price should precede one that merely describes it as expensive. Graded ranking incorporates such distinctions into the embedding space.

We compare linear gain, $\mathcal G(r)=r$, with exponential gain, $\mathcal G(r)=2^r-1$, using the shared three-grade labels. Linear gain assigns a gain difference of 1 to both adjacent boundaries. Exponential gain assigns a difference of 2 to grades 2 versus 1 and a difference of 1 to grades 1 versus 0, prioritizing fully satisfying candidates over related but insufficient ones. Both variants use the candidate-pair selection and position discounts defined in Section~\ref{sec:embedding-ranking}. Candidate-distribution distillation provides additional supervision through the teacher's relative similarity distribution over candidates.

\begin{table}[htbp]
\centering
\fontsize{8.5}{10.5}\selectfont
\setlength{\tabcolsep}{2pt}
\renewcommand{\arraystretch}{1.22}
\caption{Graded ranking and candidate-distribution distillation. The first-stage model is the starting point. KD denotes candidate-distribution distillation. All values are percentages.}
\label{tab:embedding-stage2-ablation}
\begin{tabularx}{\textwidth}{@{}>{\raggedright\arraybackslash}Xc*{5}{r}@{\hspace{5pt}}*{5}{r}@{}}
\toprule
\multirow{2}{=}{\textbf{Ranking configuration}} & \multirow{2}{*}{\textbf{KD}} & \multicolumn{5}{c}{\shortstack{\textbf{0.5M candidate pool}\\(512,655 notes)}} & \multicolumn{5}{c@{}}{\shortstack{\textbf{1.9M candidate pool}\\(1,904,054 notes)}} \\
\cmidrule(lr){3-7}\cmidrule(l){8-12}
 &  & \multicolumn{1}{c}{\retrievalmetric{Hit@1}} & \multicolumn{1}{c}{\retrievalmetric{Hit@10}} & \multicolumn{1}{c}{\retrievalmetric{Hit@100}} & \multicolumn{1}{c}{\retrievalmetric{Recall@100}} & \multicolumn{1}{c}{\retrievalmetric{NDCG@10}} & \multicolumn{1}{c}{\retrievalmetric{Hit@1}} & \multicolumn{1}{c}{\retrievalmetric{Hit@10}} & \multicolumn{1}{c}{\retrievalmetric{Hit@100}} & \multicolumn{1}{c}{\retrievalmetric{Recall@100}} & \multicolumn{1}{c}{\retrievalmetric{NDCG@10}} \\
\midrule
First-stage model & \textemdash & 80.60 & 96.86 & 99.55 & 94.73 & 79.24 & 66.09 & 92.07 & 98.57 & 87.62 & 64.84 \\
Linear gain & \textemdash & 80.84 & 96.88 & 99.54 & 95.00 & 79.67 & 66.53 & 92.12 & 98.58 & 88.24 & 65.40 \\
Exponential gain & \textemdash & 81.26 & 96.93 & 99.54 & 95.53 & 80.38 & 67.41 & 92.20 & 98.62 & 89.34 & 66.44 \\
\rowcolor{black!6}
Exponential gain & $\checkmark$ & \textbf{81.62} & \textbf{97.00} & \textbf{99.56} & \textbf{95.79} & \textbf{80.64} & \textbf{67.70} & \textbf{92.43} & \textbf{98.63} & \textbf{89.51} & \textbf{66.59} \\
\bottomrule
\end{tabularx}
\end{table}

\textbf{Graded ranking.} Exponential gain outperforms linear gain in both galleries. On the 1.9M gallery, it improves Hit@1 from 66.53\% to 67.41\% and NDCG@10 from 65.40\% to 66.44\%. The corresponding improvements on the 0.5M gallery are 0.42 and 0.71 percentage points. These results support exponential gain for improving the early placement of fully satisfying notes.

\textbf{Distribution distillation.} Adding distillation to exponential-gain ranking further increases Hit@1 to 67.70\% on the larger gallery and 81.62\% on the smaller gallery. The teacher distribution provides useful supervision beyond the ordering specified by discrete relevance grades. The second-stage configuration reaches 67.70\% Hit@1 and 89.51\% Recall@100 on the larger gallery, compared with 66.09\% and 87.62\% for the first-stage checkpoint.

\subsubsection{Query Adaptation with a Fixed Document Index}

\label{sec:embedding-eval-adaptation}

Updating document representations in an industrial retrieval system requires corpus-wide encoding and index updates. We therefore optimize the query encoder while retaining the existing document embeddings and ANN index.

Query-side LoRA adapts the fusion of the full image, reference region, and textual constraints. Query-conditioned residual projection adjusts how the resulting features map into the fixed document space. The two components are jointly optimized in the combined configuration. CSLS subsequently corrects candidate scores using local neighborhood density.

Table~\ref{tab:embedding-query-adaptation} starts from the model trained with graded ranking and distillation, then evaluates query-side LoRA, its combination with residual projection, and the additional effect of CSLS.

\begin{table}[H]
\centering
\fontsize{8.5}{10.5}\selectfont
\setlength{\tabcolsep}{2pt}
\renewcommand{\arraystretch}{1.22}
\caption{Query adaptation and candidate score correction. All configurations share the same document embeddings and index. Residual denotes the query-conditioned dual-branch residual projection; CSLS rescores the initial candidate set. All values are percentages.}
\label{tab:embedding-query-adaptation}
\begin{tabularx}{\textwidth}{@{}c>{\centering\arraybackslash}Xc*{5}{r}@{\hspace{5pt}}*{5}{r}@{}}
\toprule
\multirow{2}{*}{\textbf{LoRA}} & \multirow{2}{=}{\centering\textbf{Residual}} & \multirow{2}{*}{\textbf{CSLS}} & \multicolumn{5}{c}{\shortstack{\textbf{0.5M candidate pool}\\(512,655 notes)}} & \multicolumn{5}{c@{}}{\shortstack{\textbf{1.9M candidate pool}\\(1,904,054 notes)}} \\
\cmidrule(lr){4-8}\cmidrule(l){9-13}
 &  &  & \multicolumn{1}{c}{\retrievalmetric{Hit@1}} & \multicolumn{1}{c}{\retrievalmetric{Hit@10}} & \multicolumn{1}{c}{\retrievalmetric{Hit@100}} & \multicolumn{1}{c}{\retrievalmetric{Recall@100}} & \multicolumn{1}{c}{\retrievalmetric{NDCG@10}} & \multicolumn{1}{c}{\retrievalmetric{Hit@1}} & \multicolumn{1}{c}{\retrievalmetric{Hit@10}} & \multicolumn{1}{c}{\retrievalmetric{Hit@100}} & \multicolumn{1}{c}{\retrievalmetric{Recall@100}} & \multicolumn{1}{c}{\retrievalmetric{NDCG@10}} \\
\midrule
\textemdash & \textemdash & \textemdash & 81.62 & 97.00 & 99.56 & 95.79 & 80.64 & 67.70 & 92.43 & 98.63 & 89.51 & 66.59 \\
$\checkmark$ & \textemdash & \textemdash & 81.82 & 97.06 & 99.56 & 95.82 & 80.83 & 68.02 & 92.54 & 98.66 & 89.64 & 66.96 \\
$\checkmark$ & $\checkmark$ & \textemdash & 82.22 & 97.17 & 99.56 & 95.87 & 81.22 & 68.54 & 92.79 & 98.72 & 89.91 & 67.59 \\
\rowcolor{black!6}
$\checkmark$ & $\checkmark$ & $\checkmark$ & \textbf{82.65} & \textbf{97.37} & \textbf{99.62} & \textbf{96.15} & \textbf{81.84} & \textbf{68.84} & \textbf{93.01} & \textbf{98.85} & \textbf{90.41} & \textbf{68.13} \\
\bottomrule
\end{tabularx}
\end{table}

\Needspace{4\baselineskip}
\textbf{Query adaptation.} LoRA improves Hit@1 by 0.20 percentage points on the 0.5M gallery and 0.32 points on the 1.9M gallery. Adding residual projection provides further gains of 0.40 and 0.52 points, respectively. Refining the projection into the fixed document space remains beneficial after adapting the underlying multimodal features. On the larger gallery, joint adaptation raises Hit@1 from 67.70\% to 68.54\% and NDCG@10 from 66.59\% to 67.59\%, with unchanged document embeddings and index.

\textbf{Candidate score correction.} CSLS further increases Hit@1 to 68.84\%, Recall@100 to 90.41\%, and NDCG@10 to 68.13\% on the larger gallery. By correcting local density effects within the retrieved candidate set, it improves the placement of relevant notes after representation adaptation.

\FloatBarrier

\subsection{Intent-Conditioned Relevance Prediction}

\label{sec:relevance-eval}

\paragraph{Data and evaluation.}
The relevance student is trained on 1,705,700 query--candidate pairs labeled through teacher agreement and third-teacher adjudication using the intent-specific prompt in Section~\ref{sec:reranker}. We evaluate it on 5,977 pairs whose reference grades are assigned independently by two teacher models, with disagreements and ambiguous cases adjudicated by human annotators. The evaluation set contains 1,177, 2,574, and 2,226 pairs with reference grades 0, 1, and 2, respectively. The student generates only the final relevance grade.

\paragraph{Evaluation tasks.}
We evaluate three tasks using the same predicted and reference grades. \textbf{Three-grade relevance classification} distinguishes grades 0, 1, and 2. \textbf{Binary satisfaction classification} distinguishes candidates that satisfy the request (grade 2) from those that do not (grades 0--1). \textbf{Binary relevance classification} distinguishes related candidates (grades 1--2) from unrelated candidates (grade 0). For the two binary tasks, grades 2 and 1--2 are mapped to the positive class, respectively. We also report COPC, defined as $\overline{y}/\overline{\hat y}$ after the corresponding label mapping. Values below one indicate a higher mean predicted grade or positive rate than the reference, while values above one indicate a lower mean.

\begin{table}[htbp]
\centering\reporttablefont
\caption{Relevance-student results on 5,977 evaluation pairs. Accuracy is a percentage; COPC is the ratio of the mean reference label to the mean predicted label.}
\label{tab:relevance-results}
\begin{tabularx}{\textwidth}{@{}Yrr@{}}
\toprule
\reporttablehead{Evaluation task} & \reporttablehead{Accuracy} & \reporttablehead{COPC} \\
\midrule
Three-grade relevance classification (0 / 1 / 2) & 73.05 & 0.9922 \\
\addlinespace[2pt]
Binary satisfaction classification (2 vs. 0--1) & 82.67 & 0.8413 \\
Binary relevance classification (1--2 vs. 0) & 87.74 & 1.0823 \\
\bottomrule
\end{tabularx}
\end{table}

\paragraph{Results.}
As shown in Table~\ref{tab:relevance-results}, the student achieves 73.05\% accuracy on three-grade relevance classification, 82.67\% on binary satisfaction classification, and 87.74\% on binary relevance classification. The three-grade COPC of 0.9922 indicates that the mean predicted grade is close to the mean reference grade.

\subsection{Efficiency Analysis}

\label{sec:efficiency}

\paragraph{Parallel execution.}
The serving pipeline overlaps independent query computations. Multimodal query encoding and hierarchical intent recognition run concurrently. The query embedding is used for multimodal candidate retrieval, after which the reranker evaluates candidates using the predicted intents and the original query--candidate evidence.

\paragraph{Serving latency.} Separate service monitoring records the P50 and P90 latencies reported in Table~\ref{tab:efficiency}. The intent recognizer takes 122~ms and 146~ms, respectively, while the complete search service takes 1,070~ms and 1,340~ms. End-to-end latency is measured at the service boundary across the production pipeline. These measurements characterize the response time of the deployed system alongside the online effectiveness evaluation.

\begin{table}[htbp]
\centering\reporttablefont
\caption{Monitored P50 and P90 latencies of the principal components and the complete search service. Query encoding and intent recognition run concurrently; multimodal retrieval is followed by intent-conditioned reranking. End-to-end latency is measured across the deployed search service.}
\label{tab:efficiency}
\begin{tabularx}{\textwidth}{@{}>{\raggedright\arraybackslash}p{0.26\textwidth}Yrr@{}}
\toprule
\reporttablehead{Execution group} & \reporttablehead{Component} & \reporttablehead{P50 (ms)} & \reporttablehead{P90 (ms)} \\
\midrule
\multirow{2}{=}{\raggedright\textbf{Query processing}\\(concurrent)}
 & Multimodal query encoding & 125 & 146 \\
 & Hierarchical intent recognition & 122 & 146 \\
\midrule
\textbf{Candidate acquisition} & Multimodal retrieval & 174 & 251 \\
\midrule
\textbf{Reranking} & Intent-conditioned reranking & 342 & 468 \\
\midrule
\textbf{Complete service} & End-to-end request & 1,070 & 1,340 \\
\bottomrule
\end{tabularx}
\end{table}

\FloatBarrier
\Needspace{6\baselineskip}
\section{Conclusion}
We presented \method, an intent-driven system for real-time multimodal search in Xiaohongshu's UGC community. Its hierarchical intent taxonomy expresses the requested outcome, and intent-specific relevance criteria align data construction, embedding retrieval, and reranking. An agentic workflow acquires candidates and verifies relevance to construct supervision for contrastive learning, graded ranking, and relevance prediction. Distribution distillation further refines the embedding model, while the reranker evaluates query--candidate evidence conditioned on hierarchical intents predicted by a compact recognizer.

Controlled offline comparisons show that intent-aware supervision improves overall retrieval under matched data-construction and training conditions. In a production A/B test, the complete pipeline increases Top-4 query click-through rate by 28.17\%, query click-through rate by 18.96\%, and average time spent on search results pages per user by 12.56\% relative to the production control. Separate service monitoring records end-to-end P50 and P90 latencies of 1.07~s and 1.34~s, respectively. Together, these results demonstrate the practical value of aligning candidate retrieval and relevance assessment through explicit intent modeling in an industrial real-time search system.

\begingroup
\small
\refstepcounter{section}
\renewcommand{\refname}{\thesection.\ References}
\phantomsection
\addcontentsline{toc}{section}{\protect\numberline{\thesection}References}
\setlength{\bibsep}{0pt plus 0.5pt}
\bibliographystyle{plainnat}
\bibliography{references}

\begin{thebibliography}{33}
\providecommand{\natexlab}[1]{#1}
\providecommand{\url}[1]{\texttt{#1}}
\expandafter\ifx\csname urlstyle\endcsname\relax
  \providecommand{\doi}[1]{doi: #1}\else
  \providecommand{\doi}{doi: \begingroup \urlstyle{rm}\Url}\fi

\bibitem[Abdallah et~al.(2026)Abdallah, Mounis, Abdalla, Kasem, Senussi,
  Mahmoud, Ali, Jatowt, and Kang]{abdallah2026mmbright}
Abdelrahman Abdallah, Mohamed~Darwish Mounis, Mahmoud Abdalla,
  Mahmoud~SalahEldin Kasem, Mostafa~Farouk Senussi, Mohamed Mahmoud, Mohammed
  Ali, Adam Jatowt, and Hyun-Soo Kang.
\newblock {MM-BRIGHT}: A multi-task multimodal benchmark for
  reasoning-intensive retrieval.
\newblock arXiv:2601.09562, 2026.
\newblock URL \url{https://arxiv.org/abs/2601.09562}.

\bibitem[Asai et~al.(2023)Asai, Schick, Lewis, Chen, Izacard, Riedel,
  Hajishirzi, and Yih]{asai2023tart}
Akari Asai, Timo Schick, Patrick Lewis, Xilun Chen, Gautier Izacard, Sebastian
  Riedel, Hannaneh Hajishirzi, and Wen-tau Yih.
\newblock Task-aware retrieval with instructions.
\newblock In \emph{Findings of the Association for Computational Linguistics:
  ACL 2023}, pages 3650--3675, Toronto, Canada, 2023. Association for
  Computational Linguistics.
\newblock \doi{10.18653/v1/2023.findings-acl.225}.
\newblock URL \url{https://aclanthology.org/2023.findings-acl.225/}.

\bibitem[Borlund(2003)]{borlund2003relevance}
Pia Borlund.
\newblock The concept of relevance in {IR}.
\newblock \emph{Journal of the American Society for Information Science and
  Technology}, 54\penalty0 (10):\penalty0 913--925, 2003.
\newblock \doi{10.1002/asi.10286}.
\newblock URL \url{https://onlinelibrary.wiley.com/doi/10.1002/asi.10286}.

\bibitem[Burges(2010)]{burges2010ranknet}
Christopher J.~C. Burges.
\newblock From {RankNet} to {LambdaRank} to {LambdaMART}: An overview.
\newblock Technical Report MSR-TR-2010-82, Microsoft Research, 2010.
\newblock URL
  \url{https://www.microsoft.com/en-us/research/publication/from-ranknet-to-lambdarank-to-lambdamart-an-overview/}.

\bibitem[Chen et~al.(2025{\natexlab{a}})Chen, Wang, Yang, Zhu, Zhao, Wei, and
  Dou]{chen2025mme5}
Haonan Chen, Liang Wang, Nan Yang, Yutao Zhu, Ziliang Zhao, Furu Wei, and
  Zhicheng Dou.
\newblock {mmE5}: Improving multimodal multilingual embeddings via high-quality
  synthetic data.
\newblock In \emph{Findings of the Association for Computational Linguistics:
  ACL 2025}, pages 8254--8275, Vienna, Austria, 2025{\natexlab{a}}. Association
  for Computational Linguistics.
\newblock \doi{10.18653/v1/2025.findings-acl.433}.
\newblock URL \url{https://aclanthology.org/2025.findings-acl.433/}.

\bibitem[Chen et~al.(2026{\natexlab{a}})Chen, Li, Wang, Liu, Wang, Jiang, and
  Dou]{chen2026dme}
Haonan Chen, Chu Li, Zhicheng Wang, Yuanwei Liu, Yuanjiang Wang, Shaohua Jiang,
  and Zhicheng Dou.
\newblock {Douyin} multimodal embedding model technical report.
\newblock arXiv:2608.02148, 2026{\natexlab{a}}.
\newblock URL \url{https://arxiv.org/abs/2608.02148}.

\bibitem[Chen et~al.(2025{\natexlab{b}})Chen, Dong, Li, He, Gao, Cao, Wu, Yang,
  Xu, Hu, Ai, and Liu]{chen2025qilin}
Jia Chen, Qian Dong, Haitao Li, Xiaohui He, Yan Gao, Shaosheng Cao, Yi~Wu, Ping
  Yang, Chen Xu, Yao Hu, Qingyao Ai, and Yiqun Liu.
\newblock {Qilin}: A multimodal information retrieval dataset with {APP}-level
  user sessions.
\newblock In \emph{Proceedings of the 48th International ACM SIGIR Conference
  on Research and Development in Information Retrieval}, pages 3670--3680,
  Padua, Italy, 2025{\natexlab{b}}. ACM.
\newblock \doi{10.1145/3726302.3730279}.
\newblock URL \url{https://doi.org/10.1145/3726302.3730279}.

\bibitem[Chen et~al.(2026{\natexlab{b}})Chen, Ju, Chen, Wang, Jiao, Zhan, Li,
  Xu, Zhao, Jia, Li, Gao, Song, Lan, Zhu, and Zheng]{chen2026pailitaovl}
Lei Chen, Chen Ju, Xu~Chen, Zhicheng Wang, Yuheng Jiao, Hongfeng Zhan, Zhaoyang
  Li, Shihao Xu, Zhixiang Zhao, Tong Jia, Lin Li, Yuan Gao, Jun Song, Jinsong
  Lan, Xiaoyong Zhu, and Bo~Zheng.
\newblock {Pailitao-VL}: Unified embedding and reranker for real-time
  multi-modal industrial search.
\newblock arXiv:2602.13704, 2026{\natexlab{b}}.
\newblock URL \url{https://arxiv.org/abs/2602.13704}.

\bibitem[Chen et~al.(2026{\natexlab{c}})Chen, Liu, Wei, Wang, and
  Tang]{chen2026umer}
Libiao Chen, Xiyang Liu, Yanheng Wei, Tao Wang, and Zhenyu Tang.
\newblock {UMER}: Unifying embedding and ranking via pair-aware discriminative
  reasoning for universal multimodal retrieval.
\newblock arXiv:2608.18504, 2026{\natexlab{c}}.
\newblock URL \url{https://arxiv.org/abs/2608.18504}.

\bibitem[Conneau et~al.(2018)Conneau, Lample, Ranzato, Denoyer, and
  J{\'e}gou]{conneau2018word}
Alexis Conneau, Guillaume Lample, Marc'Aurelio Ranzato, Ludovic Denoyer, and
  Herv{\'e} J{\'e}gou.
\newblock Word translation without parallel data.
\newblock In \emph{International Conference on Learning Representations}, 2018.
\newblock URL \url{https://arxiv.org/abs/1710.04087}.

\bibitem[Deng et~al.(2025)Deng, Sun, Chen, Yang, Wang, and Song]{deng2025muka}
Lianghao Deng, Yuchong Sun, Shizhe Chen, Ning Yang, Yunfeng Wang, and Ruihua
  Song.
\newblock {MuKA}: Multimodal knowledge augmented visual information-seeking.
\newblock In \emph{Proceedings of the 31st International Conference on
  Computational Linguistics}, pages 9675--9686, Abu Dhabi, UAE, 2025.
  Association for Computational Linguistics.
\newblock URL \url{https://aclanthology.org/2025.coling-main.647/}.

\bibitem[Douze et~al.(2021)Douze, Tolias, Pizzi, Papakipos, Chanussot,
  Radenovic, Jenicek, Maximov, Leal-Taix{\'e}, Elezi, Chum, and
  Ferrer]{douze2021imagesimilarity}
Matthijs Douze, Giorgos Tolias, Ed~Pizzi, Zo{\"e} Papakipos, Lowik Chanussot,
  Filip Radenovic, Tomas Jenicek, Maxim Maximov, Laura Leal-Taix{\'e}, Ismail
  Elezi, Ond{\v{r}}ej Chum, and Cristian~Canton Ferrer.
\newblock The 2021 image similarity dataset and challenge.
\newblock arXiv:2106.09672, 2021.
\newblock URL \url{https://arxiv.org/abs/2106.09672}.

\bibitem[Hinton et~al.(2015)Hinton, Vinyals, and Dean]{hinton2015distilling}
Geoffrey Hinton, Oriol Vinyals, and Jeff Dean.
\newblock Distilling the knowledge in a neural network, 2015.
\newblock URL \url{https://arxiv.org/abs/1503.02531}.

\bibitem[Hu et~al.(2022)Hu, Shen, Wallis, Allen-Zhu, Li, Wang, Wang, and
  Chen]{hu2022lora}
Edward~J. Hu, Yelong Shen, Phillip Wallis, Zeyuan Allen-Zhu, Yuanzhi Li, Shean
  Wang, Lu~Wang, and Weizhu Chen.
\newblock {LoRA}: Low-rank adaptation of large language models.
\newblock In \emph{International Conference on Learning Representations}, 2022.
\newblock URL \url{https://arxiv.org/abs/2106.09685}.

\bibitem[J{\"a}rvelin and Kek{\"a}l{\"a}inen(2002)]{jarvelin2002cumulated}
Kalervo J{\"a}rvelin and Jaana Kek{\"a}l{\"a}inen.
\newblock Cumulated gain-based evaluation of {IR} techniques.
\newblock \emph{ACM Transactions on Information Systems}, 20\penalty0
  (4):\penalty0 422--446, 2002.
\newblock \doi{10.1145/582415.582418}.
\newblock URL \url{https://doi.org/10.1145/582415.582418}.

\bibitem[Jiang et~al.(2025)Jiang, Meng, Yang, Yavuz, Zhou, and
  Chen]{jiang2025vlm2vec}
Ziyan Jiang, Rui Meng, Xinyi Yang, Semih Yavuz, Yingbo Zhou, and Wenhu Chen.
\newblock {VLM2Vec}: Training vision-language models for massive multimodal
  embedding tasks.
\newblock In \emph{The Thirteenth International Conference on Learning
  Representations}, 2025.
\newblock URL \url{https://openreview.net/forum?id=TE0KOzWYAF}.

\bibitem[Li et~al.(2026)Li, Zhang, Long, Chen, Song, Bai, Yang, Xie, Yang, Liu,
  Zhou, and Lin]{li2026qwen3vl}
Mingxin Li, Yanzhao Zhang, Dingkun Long, Keqin Chen, Sibo Song, Shuai Bai,
  Zhibo Yang, Pengjun Xie, An~Yang, Dayiheng Liu, Jingren Zhou, and Junyang
  Lin.
\newblock {Qwen3-VL-Embedding} and {Qwen3-VL-Reranker}: A unified framework for
  state-of-the-art multimodal retrieval and ranking.
\newblock arXiv:2601.04720, 2026.
\newblock URL \url{https://arxiv.org/abs/2601.04720}.

\bibitem[Lin et~al.(2025)Lin, Lee, Shoeybi, Lin, Catanzaro, and
  Ping]{lin2025mmembed}
Sheng-Chieh Lin, Chankyu Lee, Mohammad Shoeybi, Jimmy Lin, Bryan Catanzaro, and
  Wei Ping.
\newblock {MM-Embed}: Universal multimodal retrieval with multimodal {LLM}s.
\newblock In \emph{The Thirteenth International Conference on Learning
  Representations}, 2025.
\newblock URL \url{https://openreview.net/forum?id=i45NQb2iKO}.

\bibitem[Lin et~al.(2024)Lin, Mei, Chen, and Byrne]{lin2024preflmr}
Weizhe Lin, Jingbiao Mei, Jinghong Chen, and Bill Byrne.
\newblock {PreFLMR}: Scaling up fine-grained late-interaction multi-modal
  retrievers.
\newblock In \emph{Proceedings of the 62nd Annual Meeting of the Association
  for Computational Linguistics (Volume 1: Long Papers)}, pages 5294--5316,
  Bangkok, Thailand, 2024. Association for Computational Linguistics.
\newblock \doi{10.18653/v1/2024.acl-long.289}.
\newblock URL \url{https://aclanthology.org/2024.acl-long.289/}.

\bibitem[Liu et~al.(2025)Liu, Zhang, Cai, Jiang, Hu, Yao, Wang, and
  Xie]{liu2025lamra}
Yikun Liu, Yajie Zhang, Jiayin Cai, Xiaolong Jiang, Yao Hu, Jiangchao Yao,
  Yanfeng Wang, and Weidi Xie.
\newblock {LamRA}: Large multimodal model as your advanced retrieval assistant.
\newblock In \emph{Proceedings of the IEEE/CVF Conference on Computer Vision
  and Pattern Recognition}, pages 4015--4025. IEEE, 2025.
\newblock URL
  \url{https://openaccess.thecvf.com/content/CVPR2025/html/Liu_LamRA_Large_Multimodal_Model_as_Your_Advanced_Retrieval_Assistant_CVPR_2025_paper.html}.

\bibitem[Liu et~al.(2021)Liu, Rodriguez-Opazo, Teney, and Gould]{liu2021cirr}
Zheyuan Liu, Cristian Rodriguez-Opazo, Damien Teney, and Stephen Gould.
\newblock Image retrieval on real-life images with pre-trained
  vision-and-language models.
\newblock In \emph{Proceedings of the IEEE/CVF International Conference on
  Computer Vision}, pages 2125--2134, Virtual, 2021. IEEE.
\newblock URL
  \url{https://openaccess.thecvf.com/content/ICCV2021/html/Liu_Image_Retrieval_on_Real-Life_Images_With_Pre-Trained_Vision-and-Language_Models_ICCV_2021_paper.html}.

\bibitem[Pan et~al.(2024)Pan, Li, Wang, Fei, Song, Ji, Lin, Liu, Chua, and
  Tang]{pan2024i3}
Kaihang Pan, Juncheng Li, Wenjie Wang, Hao Fei, Hongye Song, Wei Ji, Jun Lin,
  Xiaozhong Liu, Tat-Seng Chua, and Siliang Tang.
\newblock {I3}: Intent-introspective retrieval conditioned on instructions.
\newblock In \emph{Proceedings of the 47th International ACM SIGIR Conference
  on Research and Development in Information Retrieval}, pages 1839--1849,
  Washington, DC, USA, 2024. ACM.
\newblock \doi{10.1145/3626772.3657745}.
\newblock URL \url{https://doi.org/10.1145/3626772.3657745}.

\bibitem[Sarkar et~al.(2023)Sarkar, Bodla, Vasileva, Lin, Beniwal, Lu, and
  Medioni]{sarkar2023outfit}
Rohan Sarkar, Navaneeth Bodla, Mariya~I. Vasileva, Yen-Liang Lin, Anurag
  Beniwal, Alan Lu, and Gerard Medioni.
\newblock {OutfitTransformer}: Learning outfit representations for fashion
  recommendation.
\newblock In \emph{Proceedings of the IEEE/CVF Winter Conference on
  Applications of Computer Vision}, pages 3601--3609. IEEE, 2023.
\newblock URL
  \url{https://openaccess.thecvf.com/content/WACV2023/html/Sarkar_OutfitTransformer_Learning_Outfit_Representations_for_Fashion_Recommendation_WACV_2023_paper.html}.

\bibitem[Su et~al.(2025)Su, Yen, Xia, Shi, Muennighoff, Wang, Liu, Shi, Siegel,
  Tang, Sun, Yoon, Ar{\i}k, Chen, and Yu]{su2025bright}
Hongjin Su, Howard Yen, Mengzhou Xia, Weijia Shi, Niklas Muennighoff, Han-yu
  Wang, Haisu Liu, Quan Shi, Zachary~S. Siegel, Michael Tang, Ruoxi Sun,
  Jinsung Yoon, Sercan~{\"O}. Ar{\i}k, Danqi Chen, and Tao Yu.
\newblock {BRIGHT}: A realistic and challenging benchmark for
  reasoning-intensive retrieval.
\newblock In \emph{The Thirteenth International Conference on Learning
  Representations}, 2025.
\newblock URL
  \url{https://proceedings.iclr.cc/paper_files/paper/2025/hash/7a0f8055c838df8e62329a76c7c6403d-Abstract-Conference.html}.

\bibitem[Vo et~al.(2019)Vo, Jiang, Sun, Murphy, Li, Fei-Fei, and
  Hays]{vo2019composing}
Nam Vo, Lu~Jiang, Chen Sun, Kevin Murphy, Li-Jia Li, Li~Fei-Fei, and James
  Hays.
\newblock Composing text and image for image retrieval: An empirical odyssey.
\newblock In \emph{Proceedings of the IEEE/CVF Conference on Computer Vision
  and Pattern Recognition}, pages 6439--6448, Long Beach, California, USA,
  2019. IEEE.

\bibitem[Wei et~al.(2024)Wei, Chen, Chen, Hu, Zhang, Fu, Ritter, and
  Chen]{wei2024uniir}
Cong Wei, Yang Chen, Haonan Chen, Hexiang Hu, Ge~Zhang, Jie Fu, Alan Ritter,
  and Wenhu Chen.
\newblock {UniIR}: Training and benchmarking universal multimodal information
  retrievers.
\newblock In \emph{Computer Vision -- ECCV 2024}, volume 15145 of \emph{Lecture
  Notes in Computer Science}, pages 387--404. Springer, 2024.
\newblock \doi{10.1007/978-3-031-73021-4_23}.
\newblock URL
  \url{https://www.ecva.net/papers/eccv_2024/papers_ECCV/html/11927_ECCV_2024_paper.php}.

\bibitem[Weller et~al.(2025{\natexlab{a}})Weller, Chang, MacAvaney, Lo, Cohan,
  Van~Durme, Lawrie, and Soldaini]{weller2025followir}
Orion Weller, Benjamin Chang, Sean MacAvaney, Kyle Lo, Arman Cohan, Benjamin
  Van~Durme, Dawn Lawrie, and Luca Soldaini.
\newblock {FollowIR}: Evaluating and teaching information retrieval models to
  follow instructions.
\newblock In \emph{Proceedings of the 2025 Conference of the Nations of the
  Americas Chapter of the Association for Computational Linguistics: Human
  Language Technologies (Volume 1: Long Papers)}, pages 11926--11942,
  Albuquerque, New Mexico, 2025{\natexlab{a}}. Association for Computational
  Linguistics.
\newblock \doi{10.18653/v1/2025.naacl-long.597}.
\newblock URL \url{https://aclanthology.org/2025.naacl-long.597/}.

\bibitem[Weller et~al.(2025{\natexlab{b}})Weller, Van~Durme, Lawrie, Paranjape,
  Zhang, and Hessel]{weller2025promptriever}
Orion Weller, Ben Van~Durme, Dawn Lawrie, Ashwin Paranjape, Yuhao Zhang, and
  Jack Hessel.
\newblock {Promptriever}: Instruction-trained retrievers can be prompted like
  language models.
\newblock In \emph{The Thirteenth International Conference on Learning
  Representations}, 2025{\natexlab{b}}.
\newblock URL
  \url{https://proceedings.iclr.cc/paper_files/paper/2025/hash/2cefdb2c4c3274b78cd450bac35228df-Abstract-Conference.html}.

\bibitem[Wu et~al.(2021)Wu, Gao, Guo, Al-Halah, Rennie, Grauman, and
  Feris]{wu2021fashioniq}
Hui Wu, Yupeng Gao, Xiaoxiao Guo, Ziad Al-Halah, Steven Rennie, Kristen
  Grauman, and Rogerio Feris.
\newblock {Fashion IQ}: A new dataset towards retrieving images by natural
  language feedback.
\newblock In \emph{Proceedings of {CVPR}}, pages 11307--11317, Virtual, 2021.
  IEEE.

\bibitem[Zhang et~al.(2024)Zhang, Luan, Hu, Lee, Qiao, Chen, Su, and
  Chang]{zhang2024magiclens}
Kai Zhang, Yi~Luan, Hexiang Hu, Kenton Lee, Siyuan Qiao, Wenhu Chen, Yu~Su, and
  Ming-Wei Chang.
\newblock {MagicLens}: Self-supervised image retrieval with open-ended
  instructions.
\newblock In \emph{Proceedings of the 41st International Conference on Machine
  Learning}, volume 235 of \emph{Proceedings of Machine Learning Research},
  pages 59403--59420. PMLR, 2024.
\newblock URL \url{https://proceedings.mlr.press/v235/zhang24an.html}.

\bibitem[Zhang et~al.(2025)Zhang, Zhang, Xie, Li, Dai, Long, Xie, Zhang, Li,
  and Zhang]{zhang2025gme}
Xin Zhang, Yanzhao Zhang, Wen Xie, Mingxin Li, Ziqi Dai, Dingkun Long, Pengjun
  Xie, Meishan Zhang, Wenjie Li, and Min Zhang.
\newblock Bridging modalities: Improving universal multimodal retrieval by
  multimodal large language models.
\newblock In \emph{Proceedings of the IEEE/CVF Conference on Computer Vision
  and Pattern Recognition}, pages 9274--9285. IEEE, 2025.
\newblock URL
  \url{https://openaccess.thecvf.com/content/CVPR2025/html/Zhang_Bridging_Modalities_Improving_Universal_Multimodal_Retrieval_by_Multimodal_Large_Language_CVPR_2025_paper.html}.

\bibitem[Zhou et~al.(2025)Zhou, Xiong, Liu, Liu, Xiao, Wang, Zhao, Zhang, and
  Lian]{zhou2025megapairs}
Junjie Zhou, Yongping Xiong, Zheng Liu, Ze~Liu, Shitao Xiao, Yueze Wang,
  Bo~Zhao, Chen~Jason Zhang, and Defu Lian.
\newblock {MegaPairs}: Massive data synthesis for universal multimodal
  retrieval.
\newblock In \emph{Proceedings of the 63rd Annual Meeting of the Association
  for Computational Linguistics (Volume 1: Long Papers)}, pages 19076--19095,
  Vienna, Austria, 2025. Association for Computational Linguistics.
\newblock \doi{10.18653/v1/2025.acl-long.935}.
\newblock URL \url{https://aclanthology.org/2025.acl-long.935/}.

\bibitem[Zhou et~al.(2026)Zhou, Mei, Li, Wang, Rao, and Lyu]{zhou2026wemm}
Junjie Zhou, Ke~Mei, Lei Li, Tianyi Wang, Fengyun Rao, and Jing Lyu.
\newblock {WeMM-Embedding}: {WeChat} multi-modal embedding technical report.
\newblock arXiv:2608.24053, 2026.
\newblock URL \url{https://arxiv.org/abs/2608.24053}.

\end{thebibliography}
\endgroup

\clearpage
\appendix

\section{Intent Taxonomy and Relevance Criteria}
\label{app:intent-taxonomy-and-relevance}
\subsection{Intent Taxonomy}
\label{app:intent-taxonomy}
Table~\ref{tab:intents} gives the complete hierarchy and retrieval objectives introduced in Section~\ref{sec:intent-scope}. The examples in Figure~\ref{fig:motivation} instantiate these objectives; the full query determines the particular target, permitted alternatives, and explicit constraints.

\begin{table}[H]
\centering\reporttablefont
\renewcommand{\arraystretch}{1.1}
\caption{Three first-level intents and seven second-level intents. The retrieval objective defines what the query seeks; candidate satisfaction is assessed separately under the query's explicit constraints.}
\label{tab:intents}
\begin{tabularx}{\textwidth}{@{}>{\raggedright\arraybackslash\bfseries}p{0.18\textwidth}>{\raggedright\arraybackslash\bfseries}p{0.27\textwidth}Y@{}}
\toprule
\reporttablehead{First-level intent} & \reporttablehead{Second-level intent} & \reporttablehead{Retrieval objective} \\
\midrule
Visual Search & Item Search & Retrieve the same or a similar physical item, including substitutes, without explicitly requesting an attribute change. \\
\addlinespace[2pt]
Visual Search & Image Search & Retrieve an image as the result, including an original, higher-resolution version, source image, or related visual asset. \\
\addlinespace[2pt]
Visual Search & Attribute Variant Search & Retrieve a version of the reference object with a specified attribute change while preserving its other relevant characteristics. \\
\addlinespace[2pt]
Visual Search & Outfit Styling & Retrieve outfit combinations or styling examples for the reference garment or stated styling goal. \\
\midrule
Information\newline Seeking & Target-Specific\newline Information Seeking & Retrieve information that must correspond to the specified target, such as a product model, person, place, work, or problem. \\
\addlinespace[2pt]
Information\newline Seeking & Transferable\newline Information Seeking & Retrieve information that remains useful when obtained from a different but applicable object or case. \\
\midrule
Unclear Intent & Unclear Intent & Handle requests for which the image and text jointly provide no actionable search intent. Short or unfamiliar text alone does not imply this label. \\
\bottomrule
\end{tabularx}
\end{table}

\subsection{Intent-Specific Relevance Criteria}
\label{app:rubrics}
Table~\ref{tab:full-rubrics} expresses the three relevance grades under the intent definitions in Table~\ref{tab:intents}. Required identity, similarity, and visual properties are determined by the complete query. These criteria describe candidate satisfaction; they do not prescribe a universal positive or negative sampling rule for every historical data source.

\begingroup
\reporttablefont
\renewcommand{\arraystretch}{1.1}
\setlength{\LTpre}{8pt}
\setlength{\LTpost}{8pt}
\begin{xltabular}{\textwidth}{@{}>{\raggedright\arraybackslash}p{0.18\textwidth}YYY@{}}
\caption{Request-conditioned relevance criteria. Target identity is required only when the request demands it. Unclear Intent is a query-level uncertainty label, not a fourth relevance grade.}
\label{tab:full-rubrics}\\
\toprule
\reporttablehead{Second-level intent} & \reporttablehead{Grade 0:}\newline no specific relevance & \reporttablehead{Grade 1:}\newline related but insufficient & \reporttablehead{Grade 2:}\newline satisfies the request \\
\midrule
\endfirsthead
\caption[]{Request-conditioned relevance criteria (continued).}\\
\toprule
\reporttablehead{Second-level intent} & \reporttablehead{Grade 0:}\newline no specific relevance & \reporttablehead{Grade 1:}\newline related but insufficient & \reporttablehead{Grade 2:}\newline satisfies the request \\
\midrule
\endhead
\midrule
\multicolumn{4}{r@{}}{\textit{Continued on next page}}\\
\endfoot
\bottomrule
\endlastfoot
\reporttablehead{Item Search} & No specific correspondence to the requested item or its stated characteristics. & A related item fails a required target, similarity, or other explicit constraint. & The same item or an allowed alternative meets the requested characteristics and constraints. \\
\addlinespace[3pt]
\reporttablehead{Image Search} & No relevant correspondence in visual content or requested visual properties. & A related visual asset lacks required content, resolution, source evidence, or another constraint. & The image fulfills the required identity or permitted visual relation and requested form. Source claims require evidence. \\
\addlinespace[3pt]
\reporttablehead{Attribute Variant Search} & No specific relationship to the reference object beyond incidental shared properties. & The reference is related, but the requested change is absent or another required characteristic is lost. & The specified attribute changes while the other relevant characteristics and explicit constraints are preserved. \\
\addlinespace[3pt]
\reporttablehead{Outfit Styling} & No specific relationship to the reference garment or stated styling goal. & A relevant garment is shown, but the requested combination is absent or a target constraint is unmet. & The outfit or styling example meets the stated garment, combination, and other explicit requirements. \\
\addlinespace[3pt]
\reporttablehead{Target-Specific Information Seeking} & No specific relationship to the designated target or question. & A related target is incorrect, the requested information is absent, or another constraint is unmet. & Information corresponds to the specified target and actually addresses the request and its constraints. \\
\addlinespace[3pt]
\reporttablehead{Transferable Information Seeking} & No substantive correspondence to the problem or applicable information need. & Related information has unsuitable assumptions, lacks the requested outcome, or violates a constraint. & The information or method is applicable and addresses the request. The supporting object or case may differ. \\
\midrule
\reporttablehead{Unclear Intent} & \multicolumn{3}{>{\raggedright\arraybackslash}p{\dimexpr0.82\textwidth-2\tabcolsep\relax}@{}}{The joint image--text query does not establish an actionable retrieval objective. No candidate grade follows from this label alone. A general-rubric judgment requires sufficient evidence; otherwise the judgment remains unresolved.} \\
\end{xltabular}
\endgroup

\paragraph{Annotation boundaries.}
Judgments depend on the complete query and observable candidate evidence. Identity alone does not establish answer presence, authorship, or method applicability. Uncertainty about the evidence is recorded separately from a valid grade-$0$ judgment.

\FloatBarrier
\Needspace{6\baselineskip}
\section{Intent Student: Training Objective}
\label{app:intent-details}
For $N$ training queries, let $\mathbf a_n$ denote the target response for query $q_n$, and let $\mathcal T_n$ denote its supervised token positions. With $\psi$ denoting the intent student's parameters, the response-token negative log-likelihood is
\begin{equation}
\label{eq:intent-loss}
\mathcal L_{\mathrm{int}}(\psi)=-\frac{1}{N}\sum_{n=1}^{N}
 \frac{1}{|\mathcal T_n|}\sum_{j\in\mathcal T_n}
 \log p_\psi(a_{n,j}\mid q_n,\mathbf a_{n,<j}).
\end{equation}
The response lists the first-level intent label followed by the second-level label and ends with an end-of-sequence token. The loss is computed only over these response tokens; input and padding tokens are masked, and teacher explanations are excluded from the target.

\FloatBarrier
\Needspace{6\baselineskip}
\section{Relevance Student: Training Objective}
\label{app:reranker-objective}
Let $\mathcal D_{\mathrm{rr}}$ denote the training set of query--candidate pairs, their provided first- and second-level intent labels, and relevance grades obtained through the teacher annotation procedure in Section~\ref{sec:reranker}. For relevance-student parameters $\phi$, the label negative log-likelihood is
\begin{equation}
\label{eq:reranker-loss}
 \mathcal L_{\mathrm{rr}}(\phi)=
 -\sum_{\substack{(q,d,i_q^{(1)},i_q^{(2)},y)\\\in\mathcal D_{\mathrm{rr}}}}
 \log p_\phi\!\left(y\mid q,d,i_q^{(1)},i_q^{(2)}\right).
\end{equation}
The target response contains only the relevance label. Teacher explanations accompany annotation but are not included in the student response target. The intent labels provided during training serve as conditioning information. At inference, the conditioning intent labels are supplied by the recognizer, as described in Section~\ref{sec:reranker}.

\clearpage
\section{Multimodal Representation Learning: Training Details}
\label{app:embedding-details}
The retrieval student and distribution teacher have 0.8B and 2B parameters, respectively. Both use 128-dimensional retrieval representations. The distribution teacher provides training supervision; the WeMM-Embedding-2B configurations in Appendix~\ref{app:wemm-evaluation} are evaluated as external retrieval baselines. The global batch size for second-stage representation learning is 384. Table~\ref{tab:embedding-settings} summarizes the settings used in Section~\ref{sec:embedding}; they are separate from the generative-component settings reported elsewhere in this report.

\paragraph{Negative-weight schedules.}
The additional-negative weight in Equation~\eqref{eq:contrastive-q2d} is
\begin{equation}
\label{eq:embedding-negative-schedule}
 w(t)=w_{\max}\min\!\left(1,\max\!\left(0,
       \frac{t-T_n+1}{R_n}\right)\right).
\end{equation}
Here, $t$ is the zero-indexed training step, $T_n$ is the activation offset, $R_n$ is the ramp duration, and $w_{\max}$ is the maximum weight. We use $T_n=66$, $R_n=66$, and $w_{\max}=1$. The margin coefficient in Equation~\eqref{eq:embedding-stage1} is
\begin{equation}
\label{eq:embedding-margin-schedule}
 \lambda_m(t)=\lambda_m^{\max}
 \min\!\left(1,\frac{t+1}{T_m}\right),
\end{equation}
where $\lambda_m^{\max}=0.25$ is its maximum value and $T_m=99$ is its ramp duration.

\begin{table}[htbp]
\centering\reporttablefont
\caption{Training settings for multimodal representation learning and query adaptation.}
\label{tab:embedding-settings}
\begin{tabularx}{\textwidth}{@{}Yr@{}}
\toprule
\reporttablehead{Setting} & \reporttablehead{Value}\\
\midrule
\reporttablegroup{2}{Representation}
Backbone feature dimension $d_h$ & 1,024\\
Retrieval dimension $d_e$ & 128\\
\reporttablegroup{2}{Contrastive learning}
Initial first-stage temperature $\tau_1$ & 0.07\\
Hard-negative count $K_{\mathrm{hard}}$ & 30\\
Score margin $m$ & 0.05\\
\reporttablegroup{2}{Graded ranking and distillation}
Second-stage global batch size & 384\\
Contrastive / ranking / distillation temperatures & 0.07 / 0.07 / 0.10\\
Ranking weight $\lambda_{\mathrm{rank}}$ & 1.0\\
Distillation weight $\lambda_{\mathrm{KD}}$ & 0.25\\
\reporttablegroup{2}{Query adaptation}
Query expert count $K_I$ & 6\\
Residual bottleneck ranks $r_1,r_2$ & 8, 64\\
Residual coefficient $\beta$ & 0.1\\
Adapted upper fusion layers $L_a$ & 12\\
LoRA rank $r_{\mathrm L}$ / scaling $\gamma_{\mathrm L}$ & 8 / 16\\
\reporttablegroup{2}{Score correction}
CSLS neighborhood $k$ / candidate depth $K_c$ & 10 / 500\\
\bottomrule
\end{tabularx}
\end{table}

\clearpage
\section{WeMM Baseline: Evaluation Details}
\label{app:wemm-evaluation}
We evaluate WeMM-Embedding-2B~\citep{zhou2026wemm} at six supported output dimensions. Each configuration retains the first $d$ coordinates of the full 2,048-dimensional query and document embeddings and applies L2 normalization. The configurations share the same model and differ only in output dimension. Table~\ref{tab:wemm-dimensions} reports the results for 20,000 queries against the 512,655-note and 1,904,054-note galleries. The main comparison in Table~\ref{tab:embedding-overall} includes 128 dimensions to match our retrieval-vector dimension and 2,048 dimensions to retain the full WeMM representation.

\paragraph{Retrieval and metrics.}
The WeMM evaluation uses exact inner-product search. GPU matrix multiplication is used for some dimensions, with CPU FAISS verification of nearly tied candidates at the top-$K$ boundary. Hit@$K$ measures retrieval of at least one annotated positive. Recall@100 averages per-query positive coverage, as in Section~\ref{sec:embedding-eval-setup}; Micro Recall@100 aggregates retrieved and total positive relations across queries.

\begin{table}[htbp]
\centering\reporttablefont
\setlength{\tabcolsep}{4pt}
\caption{WeMM-Embedding-2B across output dimensions. All values are percentages, retaining the three-decimal precision of the recorded results. Recall@100 uses per-query averaging; Micro Recall@100 aggregates positive relations. All dimensions use the same 20,000 queries.}
\label{tab:wemm-dimensions}
\begin{tabularx}{\textwidth}{@{}Y*{7}{>{\raggedleft\arraybackslash}X}@{}}
\toprule
\textbf{Gallery} & \textbf{Dim} & \textbf{Hit@1} & \textbf{Hit@5} & \textbf{Hit@10} & \textbf{Hit@100} & \textbf{Recall@100} & \shortstack{\textbf{Micro}\\\textbf{Recall@100}}\\
\midrule
0.5M & 64 & 30.395 & 50.020 & 58.005 & 80.425 & 45.234 & 34.032\\
 & 128 & 37.945 & 58.515 & 65.810 & 85.305 & 52.428 & 41.308\\
 & 256 & 42.840 & 63.660 & 70.630 & 87.885 & 57.019 & 46.692\\
 & 512 & 44.805 & 65.525 & 72.370 & 88.530 & 58.512 & 48.603\\
 & 1,024 & 45.635 & 66.090 & 73.140 & 89.075 & 59.546 & 49.954\\
 & 2,048 & \textbf{46.140} & \textbf{66.580} & \textbf{73.585} & \textbf{89.220} & \textbf{60.081} & \textbf{50.599}\\
\midrule
1.9M & 64 & 19.560 & 36.185 & 43.910 & 68.710 & 32.047 & 22.434\\
 & 128 & 25.415 & 44.365 & 52.475 & 75.270 & 38.510 & 28.512\\
 & 256 & 29.480 & 49.530 & 57.700 & 78.900 & 42.937 & 33.156\\
 & 512 & 30.980 & 51.640 & 59.720 & 80.315 & 44.591 & 34.975\\
 & 1,024 & 31.780 & 52.300 & 60.595 & 80.840 & 45.531 & 36.135\\
 & 2,048 & \textbf{32.260} & \textbf{52.915} & \textbf{60.935} & \textbf{81.055} & \textbf{46.002} & \textbf{36.737}\\
\bottomrule
\end{tabularx}
\end{table}

\end{document}